\documentclass[10pt]{article}

\usepackage{amsmath}
\usepackage{amssymb}
\usepackage[margin=0.8in]{geometry}
\usepackage{xcolor}
\usepackage{graphicx}
\usepackage{hyphenat}
\usepackage{bm}
\usepackage{siunitx}
\usepackage{enumerate}
\usepackage{abstract}
\usepackage{float}
\usepackage{placeins}

\usepackage[superscript]{cite}

\usepackage{draftwatermark}
\SetWatermarkText{\bf PREPRINT ~~~ PREPRINT ~~~ PREPRINT   ~~~ PREPRINT}
\SetWatermarkFontSize{0.8cm}
\SetWatermarkAngle{90}
\SetWatermarkHorCenter{20.5cm}
\SetWatermarkColor[gray]{0.75}

\title{A gallery of maximum-entropy distributions: 14 and 21 moments}

\author{Stefano Boccelli$^1$\thanks{Corresponding author: stefano.boccelli@polimi.it}, 
Fabien Giroux$^1$, James G.~McDonald$^1$\\[2ex]
\textit{\normalsize $^1$University of Ottawa, ON, Canada.}
}

\begin{document}

\maketitle

\begin{abstract}
This work explores the different shapes that can be realized by the one-particle 
velocity distribution functions (VDFs) associated with the fourth-order maximum-entropy moment method.
These distributions take the form of an exponential of a polynomial of the particle velocity, with terms up to the fourth-order.
The 14- and 21-moment approximations are investigated.
Various non-equilibrium gas states are probed throughout moment space.
The resulting maximum-entropy distributions deviate strongly from the equilibrium VDF, 
and show a number of lobes and branches.
The Maxwellian and the anisotropic Gaussian distributions are recovered as special cases.
The eigenvalues associated with the maximum-entropy system of transport equations are also illustrated for some selected gas states.
Anisotropic and/or asymmetric non-equilibrium states are seen to be associated with a non-uniform spacial propagation of perturbations.
\end{abstract}

\section{Introduction}\label{sec:intro}

In the kinetic theory of gases, thermodynamic states are described, statistically, by the use of the one-particle 
distribution function \cite{ferziger1972mathematical}.
This distribution is also known as phase density, as it represents the (expected) number of particles 
in a unitary phase-space volume.
In classical mechanics, one often considers a six-dimensional phase space, composed of three spacial components, $\bm{x}$, 
and three particle velocity components, $\bm{v}$.
This results in the velocity distribution function (VDF).
We consider here a single-species monatomic gas, and write the VDF as $f = f(\bm{x}, \bm{v}, t)$.
In this work, we are concerned with investigating the different shapes assumed by the distribution function in velocity space.
Therefore, the dependence on time, $t$, and space, $\bm{x}$, is dropped. 

Under Boltzmann's H-theorem assumptions, elastic collisions bring the distribution function towards 
a Maxwellian (equilibrium) distribution \cite{cercignani1988boltzmann},
\begin{equation}\label{eq:Maxwellian-VDF-eq}
  \mathcal{M}(\bm{v}) = \frac{\rho}{m} \left( \frac{1}{2 \pi} \frac{\rho}{P} \right)^{3/2} \exp \left( - \frac{\rho}{P} \frac{(\bm{v} - \bm{u})^2}{2} \right) \, ,
\end{equation}

\noindent where $m$ is the particle mass, and $\rho$, $P$ and $\bm{u}$ are the mass density, the hydrostatic pressure 
and the bulk velocity of the gas.
Yet, non-equilibrium distributions are frequent in a variety of physical and technological fields.
For instance, non-equilibrium situations are often encountered in high Knudsen number (rarefied) flows, 
where the collisional mean free path is comparable to the physical size of the considered problem.
A typical example consists of the study of the internal structure of shock waves \cite{pham1989nonequilibrium},
whose thickness scales with the mean free path itself.
Other examples include vacuum laboratory equipment \cite{vesper2022diffusive} and
micro-scale devices \cite{harley1995gas}.
In high-Mach-number flows, one might also observe non-equilibrium VDFs, as the advection time might be comparable to the 
collisional relaxation time.
This situation is frequently encountered in hypersonic atmospheric-entry flows \cite{lofthouse2007effects}.
In the mentioned scenarios, the VDF often appears, roughly, as the superposition of multiple Maxwellians, 
each being characteristic of a different particle population.
For instance, in the shock-wave region of a hypersonic flow, one observes (i) a cold and fast free-stream population and 
(ii) a warmer and slower component associated with the post-shock gas state \cite{munafo2014spectral,boccelli2023modeling}. 

Even richer non-equilibrium situations are often encountered in plasmas, where the Lorentz force acting on 
electric charges further modifies the shape of the VDF.
For instance, in cross-field devices such as magnetrons or Hall effect thrusters \cite{boeuf2017tutorial},
the magnetic field tends to rotate the particle velocity components, and its combination with an
electric field might result in cross-field drift velocities, leading to toroidal or asymmetric VDFs \cite{shagayda2012stationary}.
Further departures from the Maxwellian are introduced by magnetic nozzles \cite{correyero2019effect}, 
chemical reactions \cite{alvarez2022regularized}, 
plasma-surface interaction processes \cite{kaganovich2007kinetic}, and kinetic 
instabilities \cite{taccogna2019latest,tarasov2021evolution}. 
Other non-equilibrium situations can be encountered in various branches of plasma physics, including 
space physics \cite{iii2022need,beth2022cometary} and fusion reactors \cite{power2021ion}.

The time and space evolution of the one-particle distribution function can be studied by solving a kinetic equation,
such as the Boltzmann or the Vlasov equations \cite{liboff2003kinetic}, either by a direct discretization 
of phase-space \cite{mieussens2000discrete} or by employing particle methods \cite{bird1994molecular,birdsall2018plasma}.
However, this approach is often computationally demanding.
This has led to the development of a class of methods, known as moment methods, where one writes evolution 
equations, not for the VDF itself, but for a finite set of its statistical moments \cite{struchtrup2005macroscopic}.
Examples of moment methods include the Grad method \cite{grad1949kinetic}, quadrature-based moment approaches
\cite{fox2009higher}, and the maximum-entropy methods \cite{muller1993extended,levermore1996moment}.
The different families of moment methods are based on assuming that the distribution function has a certain shape.
For instance, in the Grad method, the VDF is approximated by a local Maxwellian, modified by a truncated series of 
Hermite polynomials.
Instead, in the maximum-entropy method, one writes the VDF as the exponential of a polynomial of the particle velocity.
This expression is general enough to produce a wide variety of shapes that can mimic the VDFs encountered in 
the aforementioned non-equilibrium situations.

The aim of this work is to investigate the variety of shapes that can be attained by two 
members of the maximum-entropy family of methods, namely the 14- and the 21-moment methods. 
In Section~\ref{sec:maxent}, we discuss the maximum-entropy framework, the thermodynamic 
constraints associated with the positivity of the distribution function (Hamburger moment 
problem), and the region of singularity known as the Junk subspace.
In Section~\ref{sec:14mom-maxent}, we introduce the 14-moment method,
and investigate the shapes that can be assumed by the associated VDF.
In Section~\ref{sec:21mom-maxent}, we then consider the 21-moment method, 
and study a set of additional shapes and topologies that can be obtained by this increased 
the set of moments.
Finally, in Section~\ref{sec:eigenvalues}, we analyze the eigenvalues (wave speeds) of the flux Jacobian
of the maximum-entropy system of equations, for some selected states, and superimpose them to the 
respective VDF plots in velocity space.

Throughout the work, we aim to providing an intuitive feeling for the way the different 
moments affect the resulting maximum-entropy distribution function (in particular, the pressure anisotropy,
the heat flux tensor components and the fourth-order moment of the VDF).


\section{The maximum-entropy (ME) method}\label{sec:maxent}

In moment methods, the state of a gas is specified through a finite set of macroscopic properties, collected in 
a state vector, $\bm{U}$.
These properties often include the density, the average momentum and the gas energy.
However, additional macroscopic quantities, such as the full pressure tensor and the heat flux, can also be employed.

Statistically, macroscopic variables are obtained as weighted averages (moments) of the distribution function, $f$.
In particular, the moment, $M_\phi$, associated with a desired particle quantity, $\phi$, is 
\begin{equation}\label{eq:moment-definition-integral-Mphi}
  M_\phi = \iiint_{-\infty}^{+\infty} \phi (\bm{v}) \, f(\bm{v}) \, \mathrm{d} \bm{v} 
         \equiv \left< \phi (\bm{v}) \, f(\bm{v}) \right> \, ,
\end{equation}

\noindent where $\bm{v}$ is the particle velocity.
Often, the peculiar velocity, $\bm{c} = \bm{v} - \bm{u}$, is employed.
For instance, the choices $\phi = m$, $m\bm{v}$ and $m c_i c_j$, result in the gas density,
average momentum and pressure tensor respectively,
\begin{equation}
  \rho = \left< m \, f \right> \ \ , \ \ \ \rho u_i = \left< m v_i \, f \right> \ \ , \ \ \ P_{ij} = \left< m c_i c_j \, f \right> \, .
\end{equation}

\noindent Here, we employ index notation, and repeated indices imply summation.
For instance, the hydrostatic pressure is $P=P_{ii}/3$.
Higher-order moments such as the heat flux tensor, $Q_{ijk}$, and the fourth-order moment, $R_{ijkl}$, 
can be obtained as
\begin{equation}\label{eq:kinetic-definition-Q-R}
  Q_{ijk} = \left< m c_i c_j c_k \, f \right> \ \ , \ \ \ R_{ijkl} = \left< m c_i c_j c_k c_l \, f \right> \, .
\end{equation}

It is often convenient to define dimensionless moments.
These are obtained by rescaling the moments by the gas density and by suitable powers of the characteristic thermal
velocity, $\sqrt{P/\rho}$.
Dimensionless moments are indicated by a superscript, $^\star$, and read
\begin{equation}\label{eq:dimensionless-moments-definition}
  \begin{cases}
    \rho^\star = 1 \, , \\
    P_{ij}^\star = P_{ij}/P \, , \\
    Q_{ijk}^\star = Q_{ijk}/\left[ \rho \left( P/\rho \right)^{3/2} \right] \, , \\
    R_{ijkl}^\star = R_{ijkl}/\left[ \rho \left( P/\rho \right)^{2} \right] \, . \\
  \end{cases}
\end{equation}

\noindent The particle velocity itself can be non-dimensionalized, following
\begin{equation}\label{eq:vel-axis-dimensionless}
  v_i^\star = v_i/\sqrt{P/\rho} \, .
\end{equation}

Notice that a state vector, $\bm{U}$, composed of a finite number of moments, 
contains only a limited amount of information, which is not sufficient to determine uniquely the original distribution function.
Indeed, there are usually multiple VDFs that lead to the same moment state.
In maximum-entropy (ME) moment methods, one selects, among all possible VDFs, the distribution that maximizes the 
statistical entropy, $S$, while being compatible with the moments, $\bm{U}$.
For a single-component classical gas, the entropy is written as 
\begin{equation}
  S = - k_B \iiint_{-\infty}^{+\infty} f \ln\frac{f}{y} \, \mathrm{d} \bm{v} \, ,
\end{equation}

\noindent where $k_B$ is the Boltzmann constant and $y$ is a scaling parameter.
For a discussion of entropy maximization in the context of gases that include quantum effects, 
the reader is referred to \cite{dreyer1987maximisation}.
It can be shown that the entropy is maximized when the VDF takes the form \cite{levermore1996moment}
\begin{equation}\label{eq:f-max-ent-exponential-polynomial}
  f^{\mathrm{ME}} = \exp\left( \bm{\alpha}^\intercal \bm{\Phi}(\bm{v}) \right) \, ,
\end{equation}

\noindent where $\bm{\alpha}\in \mathbb{R}^N$ is a vector of coefficients, while the vector $\bm{\Phi}(\bm{v})$ collects the $N$ functions, $\phi(\bm{v})$, employed to 
generate the set of moments of interest, $\bm{U}$.
The maximum power of the particle velocity, appearing in $\bm{\Phi}(\bm{v})$, defines the order of the method.
The Maxwellian distribution of Eq.~\eqref{eq:Maxwellian-VDF-eq} can be shown to be a second-order ME distribution.
Its anisotropic version, the Gaussian distribution, is also a second-order ME VDF, reading
\begin{equation}\label{eq:Gaussian-VDF-eq}
  \mathcal{G} = \frac{n}{(2 \pi)^{3/2} \left[\det \left(\Theta_{ij}\right) \right]^{1/2}}  
                \exp \left( - \frac{1}{2} \Theta_{ij}^{-1} c_i c_j \right)\, ,
\end{equation}

\noindent where $n=\rho/m$ is the number density, and $\Theta_{ij} = P_{ij}/\rho$.
In this work, we consider two fourth-order members of the ME family of moment methods,
namely the 14- and 21-moment models.
In these models, the VDF is written as a function of 14 and 21 parameters, $\bm{\alpha}$, 
which are ultimately connected to the 14 or 21 entries in the state vector, $\bm{U}$.
In these methods, the particle velocity vector, $\bm{\Phi}(\bm{v})$, reads
\begin{equation}\label{eq:m-fourth-order-14-21}
  \begin{aligned}
    \bm{\Phi}_{14} &= \left(1, v_i, v_i v_j, v_i v^2, v^4  \right) \, , \\
    \bm{\Phi}_{21} &= \left(1, v_i, v_i v_j, v_i v_j v_k, v^4  \right) \, . \\
  \end{aligned}
\end{equation}

\noindent The resulting distribution functions are discussed in Sections~\ref{sec:14mom-maxent} 
and \ref{sec:21mom-maxent}.
These VDFs include the Maxwellian (Eq.~\ref{eq:Maxwellian-VDF-eq}) and the anisotropic Gaussian distribution (Eq.~\ref{eq:Gaussian-VDF-eq}) 
as special cases, but can result in large deviations from equilibrium in the general case,
introducing asymmetries and a non-Maxwellian kurtosis (heavier/thinner tails and holes in the VDF).
Other fourth-order ME models are the 26- and the 35-moment methods \cite{levermore1996moment}, 
which are not considered in this work.

Ultimately, the entropy maximization is a constrained optimization problem, where one needs to find 
the 14 or 21 coefficients, $\bm{\alpha}$, under the constraint that the resulting moments are equal to $\bm{U}$.
To date, no general analytical solution is available for this problem, and we thus employ a numerical solutions,
as discussed in Section~\ref{sec:numerical-entropy-maximization}.


\subsection{Physical realizability boundary and Junk subspace}\label{sec:phys-real-boundary-and-junk}

The distribution function, $f$, represents the expected number of particles in a phase-space volume, and 
is thus a non-negative quantity.
This has a number of implications.
For instance, the density is necessarily non-negative.
The same stands for every even-order moment of the VDF, such as the gas energy, 
or the contracted fourth-order moment, $R_{iijj}$, defined in Eq.~\eqref{eq:kinetic-definition-Q-R}.
Additional non-trivial relations between the moments exist, and can be obtained by solving the 
Hamburger moment problem \cite{hamburger1944hermitian}.
In particular, it is possible to show that $R_{iijj}$ has a lower threshold that depends on the pressure tensor and on
the heat flux \cite{mcdonald2013affordable}.
In non-dimensional form,
\begin{equation}\label{eq:realizability-14mom-Riijj-dimensionless}
  R_{iijj}^\star \ge Q_{ikk}^\star(P^{-1})^\star_{ij}Q^\star_{jll} + 9 \, .
\end{equation}

\begin{figure}[h]%
\centering
\includegraphics[width=0.5\textwidth]{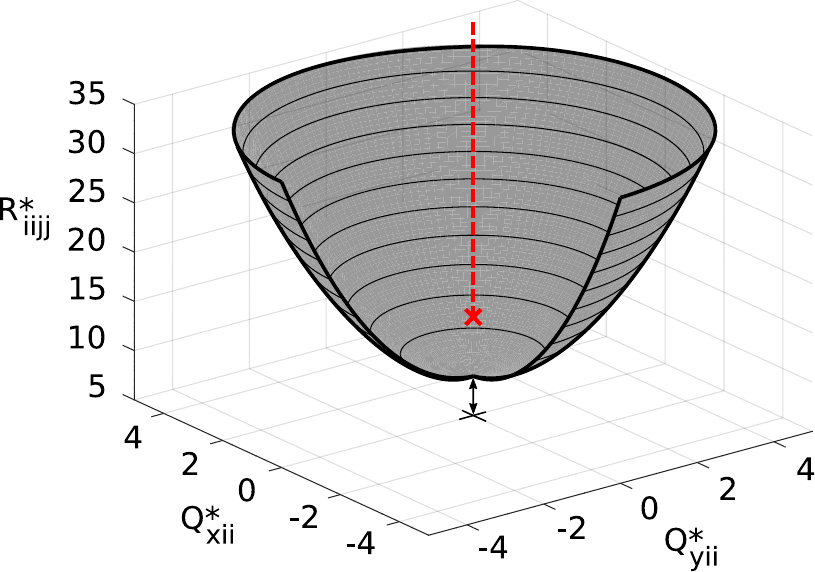}
\caption{Physical realizability boundary (paraboloid, Eq.~\eqref{eq:realizability-14mom-Riijj-dimensionless}) and Junk subspace (red dashed line, 
         Eq.~\eqref{eq:Junk-subspace-14mom}) for the case of an isotropic gas, with $P_{ij} = P\delta_{ij}$, 
         and for $Q_{zii}=0$.
         The minimum of the paraboloid is at location $Q^\star_{xii}=Q^\star_{yii} = 0$ and $R^\star_{iijj}=9$.
         The realizability paraboloid has been cropped to permit the visualization of the equilibrium point (red cross),
         situated at $Q^\star_{xii}=Q^\star_{yii} = 0$ and $R^\star_{iijj}=15$.}
\label{fig:realizability-plot-3D-gray}
\end{figure}

\noindent For an extension of this result to special-relativistic gases, see \cite{boccelli2022realizability}.
No positive distribution function can result in a gas state that lies below this threshold.
The equality sign in Eq.~\ref{eq:realizability-14mom-Riijj-dimensionless} defines a hypersurface in moment space that is referred to as the physical realizablity boundary.
Visualizing this boundary is not trivial due to the significant number of moments involved.
In Figure~\ref{fig:realizability-plot-3D-gray}, we propose a visualization, by assuming an isotropic
pressure state, $P_{ij}^\star = \delta_{ij}$, where $\delta_{ij}$ is the Kronecker delta, and 
by assuming the heat flux in one direction to be zero, $Q_{zii}^\star = 0$.
In the remaining two-dimensional space, the realizability boundary is a paraboloid.
The minimum allowed value for the fourth-order moment is $R_{iijj}^\star = 9$, and is realized when the 
heat flux is zero.
Whenever a non-zero heat flux is present, $R_{iijj}^\star$ must also increase.

\noindent Notice that, in three-dimensional velocity space, the Hamburger method gives a necessary but not sufficient condition.
Physically realizable points are located above the paraboloid (see Fig.~\ref{fig:realizability-plot-3D-gray}).
Each point is associated with a different maximum-entropy VDF,
except for a zero-measure subspace, where the entropy maximization problem has no solution.
This is known as the Junk subspace \cite{junk1998domain} and is represented as a red dashed 
line in Fig.~\ref{fig:realizability-plot-3D-gray},
\begin{equation}\label{eq:Junk-subspace-14mom}
  Q_{ijj} = 0 \ \ , \ \ \ R_{iijj} \ge \frac{2 P_{ji} P_{ij} + P_{ii} P_{jj}}{\rho} \, .
\end{equation}

\noindent For an isotropic pressure tensor, the bottom of the Junk subspace corresponds to the moments
$Q_{ijj} = 0$ and $R_{iijj} = 15 P^2/\rho$ (or equivalently $R_{iijj}^\star = 15$).
These happen to be the moments that characterize a Maxwellian distribution, 
\begin{equation}
  Q_{ijj}^{\mathcal{M}}=0 \ \ , \ \ \  R_{iijj}^{\mathcal{M}}=15 P^2/\rho \, .
\end{equation}

\noindent As one introduces some pressure anisotropy, the Junk subspace raises.
Its bottom is located at a super-Maxwellian value of the fourth-order moment, which coincides with that of a Gaussian distribution 
(Eq.~\eqref{eq:Gaussian-VDF-eq}),
\begin{equation}
  Q_{ijj}^{\mathcal{G}}=0 \ \ , \ \ \  R_{iijj}^\mathcal{G} = \frac{2 P_{ji} P_{ij} + P_{ii} P_{jj}}{\rho} \ge R_{iijj}^{\mathcal{M}} \, .
\end{equation}

\begin{figure}[h]%
\centering
\includegraphics[width=0.5\textwidth]{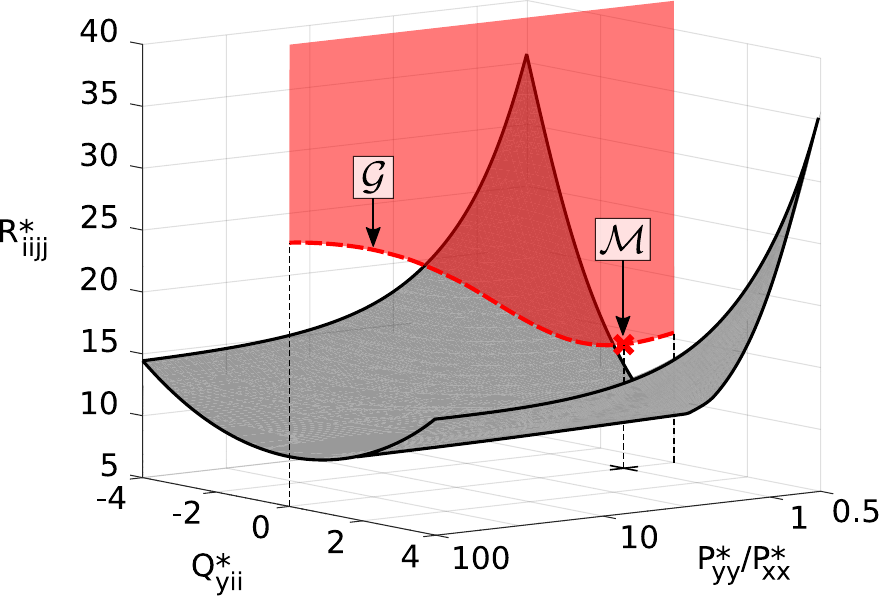}
\caption{Physical realizability boundary from Eq.~\eqref{eq:realizability-14mom-Riijj-dimensionless} and Junk subspace (red dashed line and 
         red transparent surface, from Eq.~\eqref{eq:Junk-subspace-14mom}) in the presence of pressure anisotropy, with $P_{xx}=P_{zz} \neq P_{yy}$. 
         Here, only the $Q_{yii}$ component of the heat flux non-zero.
         The Maxwellian distribution, $\mathcal{M}$, is located at the lowest point of the Junk subspace (red symbol), for $P_{yy}^\star=1$
         (isotropic state).
         For anisotropic states, the lower boundary of the Junk subspace corresponds to a Gaussian distribution, $\mathcal{G}$.}
\label{fig:realizability-plot-3D-Qyii-Pyy}
\end{figure}

\noindent This is shown in Fig.~\ref{fig:realizability-plot-3D-Qyii-Pyy}, where we show the physical realizability boundary 
and the Junk subspace in the presence of anisotropy.

The ME VDFs take some noteworthy shapes at specific points in moment space.
First of all, in absence of anisotropy, the ME VDFs recover the Maxwellian distribution at the bottom of the Junk subspace.
Indeed, for this specific combination of moments, the Maxwellian is the distribution that guarantees the largest entropy.
Along the remaining points at the lower end of the Junk subspace, the ME VDFs recover the Gaussian distribution.
Finally, as one approaches the physical realizability boundary, the ME VDFs can be seen to condense (although non-uniformly) 
on a spherical crust in velocity 
space \cite{mcdonald2013affordable}, with radius $\Upsilon$ and centred on the velocity $\hat{u}_i$,
\begin{equation}\label{eq:radius-sphere-velspace-14mom}
  \begin{cases}
    \Upsilon = \sqrt{\hat{u}_i \hat{u}_i + P_{ii}/\rho } \, , \\
    \hat{u}_i = \tfrac{1}{2} (P^{-1})_{ij} Q_{jkk} \, .
  \end{cases}
\end{equation}


\subsection{Numerical solution of the entropy maximization problem}\label{sec:numerical-entropy-maximization}

For a maximum-entropy distribution, the problem of finding the coefficients, 
$\bm{\alpha}$, whose VDF integrates to a set of target moments, $\bm{U}$, can be written as
\begin{equation}\label{eq:numerical-procedure-find-the-moments}
  \bm{U} - \iiint_{-\infty}^{+\infty} \bm{\Phi}(\bm{v}) \, \exp\left(\bm{\alpha}^\intercal \bm{\Phi}(\bm{v}) \right) \, \mathrm{d} \bm{v} = H(\bm{\alpha}) = 0 \, .
\end{equation}

For maximum-entropy distributions of order higher than two, no analytical solution of this problem is available.
In this work, we adopt a numerical approach, and solve the problem $H(\bm{\alpha}) = 0$ using Newton iterations.
The integration domain is first divided in a number of cells, and the integral is evaluated inside each cell using 
a 5-point Gaussian quadrature method.
While evaluating the integral in Eq.~\eqref{eq:numerical-procedure-find-the-moments}, it is necessary to ensure that the numerical domain is 
sufficiently large to include the whole VDF.
In certain conditions---particularly, close to the Junk subspace, as shown in Section~\ref{sec:14-mom-Junk-subspace-VDFs}---the VDF 
typically grows significantly in size, and one might need to employ some adjustment of the domain.
If the integration domain is too small, it is possible that the optimizer returns a set of parameters, $\bm{\alpha}$, 
that satisfy Eq.~\ref{eq:numerical-procedure-find-the-moments} on the compact numerical domain, but that lead to a diverging VDF 
in the actual $(-\infty,+\infty)$ region.
In our optimizer, the domain size is checked after each optimization round and enlarged dynamically if necessary.
Also, to check that the number of velocity points employed in the integration is sufficient to ensure accuracy, 
the moments of the resulting VDF are re-computed after each optimization, and are compared to the initial target values.

For most of the VDFs shown in this work, the unknowns, $\alpha_i$, were initialized at their Maxwellian values.
However, strongly non-equilibrium states located near the Junk subspace might require that one adopts an incremental approach:
instead of going directly from equilibrium to the final target point, one can define a set of intermediate targets, and 
approach the final desired state incrementally.
This is particularly important, for instance, for gas states located close to the Junk subspace.

To aid the convergence of the Newton iterations, it is advised that the algorithm is fed with dimensionless moment states, 
as defined in Eq.~\ref{eq:dimensionless-moments-definition}, and with a zero bulk velocity.
By a rigid translation and a rescaling of the velocity axes, the results are easily transported back to the original 
dimensional target moments.
Other transformations that help convergence are possible.
For instance, rather than a uniform scaling (Eq.~\ref{eq:vel-axis-dimensionless}),
one can both rotate and stretch the velocity axes as to make the whole pressure tensor diagonal and unitary.
This transformation standardizes all moments, up to the second order, and the description of a non-equilibrium state is then left 
to the heat-flux tensor and to the fourth-order moment.
For further details, the reader is referred to \cite{abramov2010multidimensional}.
However, this transformation is not employed or further discussed in this work.


\section{The 14-moment maximum-entropy moment method}\label{sec:14mom-maxent}

In this section, we investigate the shapes that can be assumed by the 14-moment ME distribution,
as a function of its moments.
This VDF reads
\begin{equation}\label{eq:14mom-VDF-expdefinition}
  f^{\mathrm{ME}}_{14} = \exp\left( \alpha_0 + \alpha_i v_i + \alpha_{ij} v_i v_j + \alpha_{i,3} v_i v^2 + \alpha_{4} v^4 \right) \, .
\end{equation}

\noindent As mentioned, this distribution can recover the Maxwellian and the anisotropic Gaussian as special cases,
but is also able to assume strongly asymmetric shapes, and can reproduce bi-modals \cite{boccelli2023modeling}.
The vector of the 14 moments of interest, $\bm{U}_{14}$, which we employ in the numerical optimization, is \cite{mcdonald2013affordable}
\begin{equation}
  \bm{U}_{14} 
  = 
    \begin{pmatrix} 
       \rho  \\ 
       \rho u_i  \\ 
       \rho u_i u_j + P_{ij} \\
       \rho u_i u_j u_j + u_i P_{jj} + 2 u_j P_{ij} + Q_{ijj} \\
       \rho u_i u_i u_j u_j + 2 u_i u_i P_{jj} + 4 u_i u_j P_{ij} + 4 u_i Q_{ijj} + R_{iijj} 
    \end{pmatrix} \, ,
\end{equation}

\noindent and the corresponding vector of primitive variables of interest, $\bm{W}_{14}$, is
\begin{equation}
  \bm{W}_{14} = \left(\rho, u_i, P_{ij}, Q_{ijj}, R_{iijj} \right)\, .
\end{equation}

\noindent It is important to remark that the 14-moment model does not depend on the full heat flux tensor, $Q_{ijk}$, but only
on its contraction, $Q_{ijj}$. 
We refer to it as the heat flux vector.
Notice that the quantity, $Q_{ijj}$, differs from the typical fluid-dynamic definition of the heat flux vector, $q_i$, by a factor 1/2, and $q_i = \tfrac{1}{2}Q_{ijj}$.
The individual elements of the full heat flux tensor cannot be selected freely, but are the result of the entropy maximization problem.
In the same fashion, the fourth-order tensor, $R_{ijkl}$, only enters the 14-moment formulation through its scalar contraction, $R_{iijj}$.

In this work, we consider a frame of reference where the bulk velocity is zero.
This has no effect on our analysis due to Galileian invariance.
Moreover, we consider dimensionless moments,
\begin{equation}
  \bm{W}_{14}^\star = \left(\rho^\star = 1, u_i^\star = 0, P_{ij}^\star, Q_{ijj}^\star, R_{iijj}^\star \right)\, .
\end{equation}

\noindent The resulting VDFs have a zero bulk velocity, and the particle velocity axes 
can be re-scaled following Eq.~\eqref{eq:vel-axis-dimensionless}.
The investigation is carried as follows:
first, starting from a Maxwellian distribution, we investigate the effect of increasing, individually, the heat flux, 
the fourth-order moment, and the degree of pressure anisotropy. 
Then, starting from these baseline states, we study the combined effect of additional non-equilibrium moments.


\subsection{Heat flux and fourth-order moment}\label{sec:heat-flux-Riijj-effect-14mom}

We start by investigating the effect of $R_{iijj}^\star$, or $Q_{ijk}^\star$, to an otherwise Maxwellian distribution.
All moments are set to their Maxwellian value, and only a single moment at a time is modified.

\begin{figure}[h]%
\centering
\includegraphics[width=0.9\textwidth]{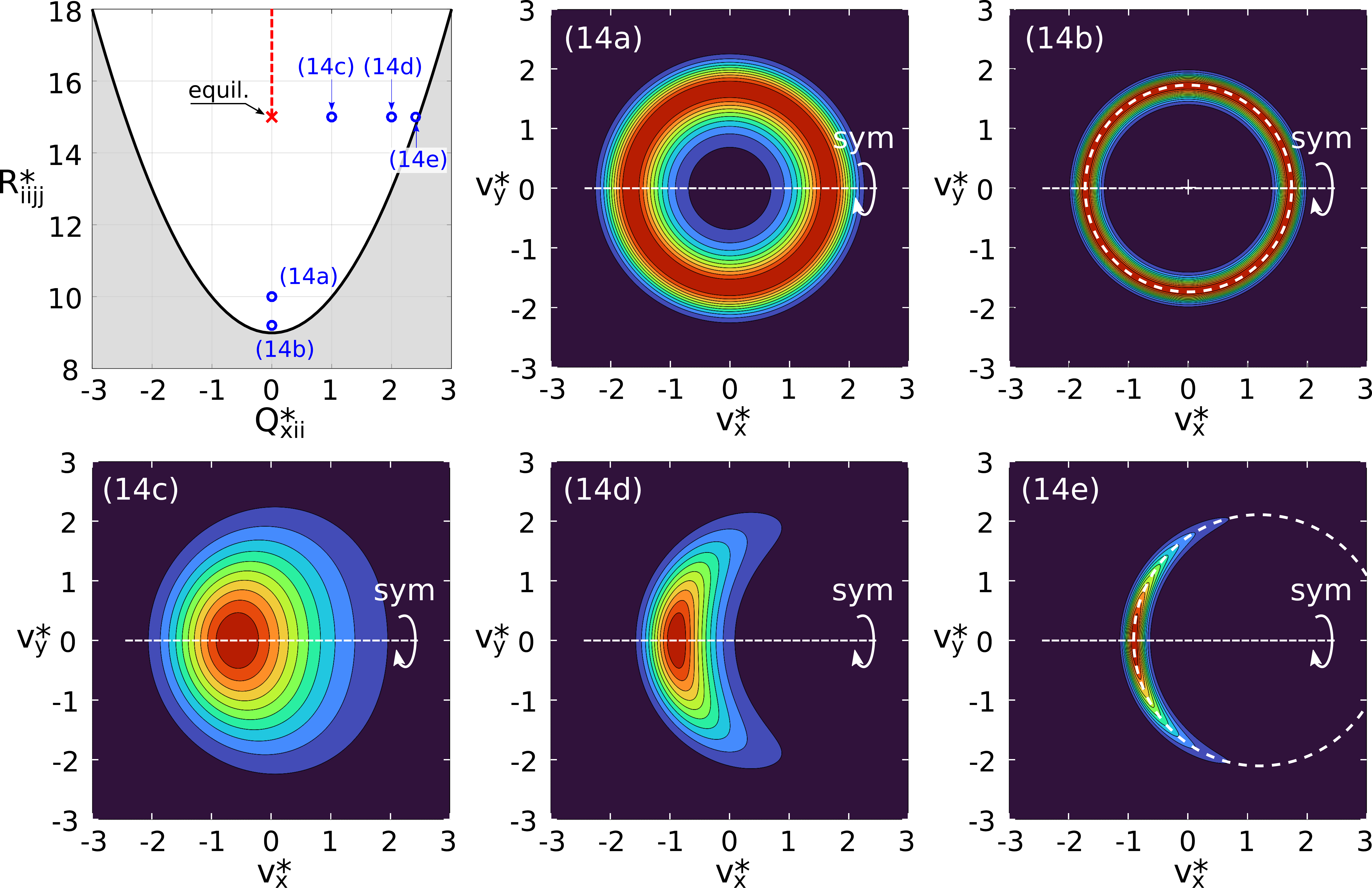}
\caption{Top-Left: dimensionless moment space; the black parabola represents the physical realizability boundary, and 
         the vertical dashed red line is the Junk subspace.
         Other panels: 
         case (14a) $R_{iijj}^\star = 10$; 
         (14b) $R_{iijj}^\star = 9.2$; 
         (14c) $Q_{xii}^\star = 1$; 
         (14d) $Q_{xii}^\star = 2$; 
         (14e) $Q_{xii}^\star = 2.4$. 
         As the physical realizability boundary is approached, the VDF collapses on a sphere (white dashed line in cases (14b) and (14e)),
         with radius from Eq.~\eqref{eq:radius-sphere-velspace-14mom}.
         The coloring of each distribution is rescaled for a better representation.}
\label{fig:14mom-R-Q-VDFs}
\end{figure}

The selected states are shown in Fig.~\ref{fig:14mom-R-Q-VDFs}-Top-Left, as points in the dimensionless moment space.
These 14-moment states are 
\begin{description}
  \item[\rm(14a)] $P_{ij}^\star = \delta_{ij} \ \ , \ \ \  Q_{ijj}^\star = 0 \ \ , \ \ \ R_{iijj}^\star = 10 \ $; 
  \item[\rm(14b)] $P_{ij}^\star = \delta_{ij} \ \ , \ \ \  Q_{ijj}^\star = 0 \ \ , \ \ \ R_{iijj}^\star = 9.2 \ $; 
  \item[\rm(14c)] $P_{ij}^\star = \delta_{ij} \ \ , \ \ \  Q_{xii}^\star = 1 \ \ , \ \ \   Q_{yii}^\star = Q_{zii}^\star = 0 \ \ , \ \ \ R_{iijj}^\star = 15 \ $;
  \item[\rm(14d)] $P_{ij}^\star = \delta_{ij} \ \ , \ \ \  Q_{xii}^\star = 2 \ \ , \ \ \   Q_{yii}^\star = Q_{zii}^\star = 0 \ \ , \ \ \ R_{iijj}^\star = 15 \ $;
  \item[\rm(14e)] $P_{ij}^\star = \delta_{ij} \ \ , \ \ \  Q_{xii}^\star = 2.4 \ \ , \ \ \ Q_{yii}^\star = Q_{zii}^\star = 0 \ \ , \ \ \ R_{iijj}^\star = 15 \ $.
\end{description}

In the states (14a) and (14b), all moments are Maxwellian, except for the contracted fourth-order moment, 
$R_{iijj}^\star < 15$.
This has the effect of modifying both the tails and the bulk of the VDF. 
In particular, a low value of $R_{iijj}^\star$ is associated with fewer particles in the tails of the distribution.
However, since the hydrostatic pressure (and thus the temperature) is fixed to $P^\star = 1$, particles also need to be 
displaced from the bulk of the distribution (centred at zero velocity), towards higher velocities.
If $R_{iijj}^\star$ is sufficiently small, the final result is a spherically symmetric VDF showing a hole in the centre.
This is shown in Fig.~\ref{fig:14mom-R-Q-VDFs}-Top-Centre and Top-Right.

\begin{figure}[h]%
\centering
\includegraphics[width=0.9\textwidth]{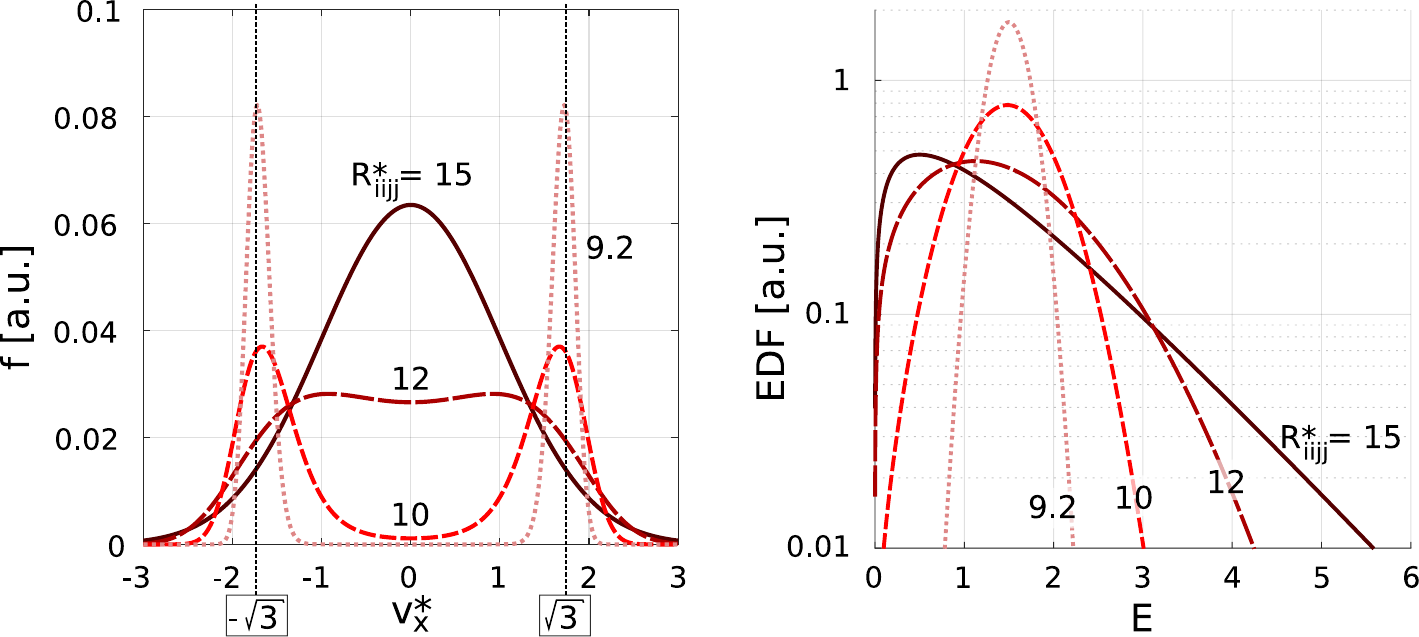}
\caption{14-moment distribution functions obtained for different values of $R_{iijj}^\star$, 
  with $Q_{ijj}^\star = 0$ and $P_{ij}^\star = \delta_{ij}$.
  Left: slices of the three-dimensional VDFs, extracted along the line $v_y^\star = v_z^\star = 0$.
  Right: EDFs, obtained by numerical evaluation of Eq.~\eqref{eq:integral-fE}.
  The cases $R_{iijj}^\star=10$ and $R_{iijj}^\star=9.2$ correspond to the VDFs (14a) and (14b) of Fig.~\ref{fig:14mom-R-Q-VDFs}.
  The two vertical dashed lines at $v_x^\star = \pm \sqrt{3}$ represent the radius of the sphere from 
  Eq.~\eqref{eq:radius-sphere-velspace-14mom}. 
  The energy axis in the Right panel assumes a unitary particle mass, $m = 1$.}
\label{fig:14mom-f-various-R}
\end{figure}

Physical realizability requires that, if the heat flux is zero, then $R_{iijj}^\star \ge 9$.
State (14b) is selected as to approach this boundary.
As discussed, close to the realizability boundary, the VDF collapses on a spherical crust, whose radius and centre are 
given by Eq.~\eqref{eq:radius-sphere-velspace-14mom}.
As the state is completely symmetric, the distribution is uniform on this sphere.
The evolution of a Maxwellian into a spherical distribution, as $R_{iijj}^\star$ is lowered, 
is also analyzed in Fig.~\ref{fig:14mom-f-various-R}.
The energy distribution function (EDF) shown in Fig.~\ref{fig:14mom-f-various-R}-Right is computed as discussed in 
Appendix~\ref{sec:appendix-EDF}. 

Cases (14c), (14d) and (14e) are obtained by moving from equilibrium towards the physical realizability boundary, keeping 
an equilibrium value for the fourth-order moment, $R_{iijj}^\star = 15$.
The heat flux can be seen to be associated with an asymmetric VDF.
Initially, this appears as a displacement of the bulk of the VDF towards one of the tails (negative velocity tail, for a 
positive heat flux), as in case (14d).
As higher values of $Q_{xjj}^\star$ are considered, the VDF bends towards its sides, and eventually collapses on a sphere.


\subsection{Pressure anisotropy}\label{sec:14mom-Pij-anisotropy}
 
We consider here the effect of the pressure tensor, $P_{ij}$, on the shape of the 14-moment ME VDFs.
First, it should be noted that, by a suitable rotation of the reference frame, it is always possible to 
write the pressure tensor in a diagonal form.
Indeed, off-diagonal components (shear terms) do not deform the VDF, and only introduce a rigid rotation 
in velocity space.
Also, the effect of the hydrostatic pressure is merely that of scaling the velocity axes.
In dimensionless form, we can thus write
\begin{equation}
  P_{ij}^\star =  \begin{bmatrix}
             P_{xx}^\star & 0            & 0 \\
             0            & P_{yy}^\star & 0 \\
             0            & 0            & P_{zz}^\star \\
           \end{bmatrix} \, .
\end{equation}

\noindent If any of the (dimensionless) principal components differs from one, the gas state is characterised by pressure (temperature) anisotropy.
To investigate this, we consider first the case where $P_{yy}^\star > P_{xx}^\star = P_{zz}^\star$.
This results in an elongated VDF along the $y$-velocity axis.
The heat flux is here set to $Q_{ijj}^\star = 0$.

\begin{figure}[h]
\centering
\includegraphics[width=0.9\textwidth]{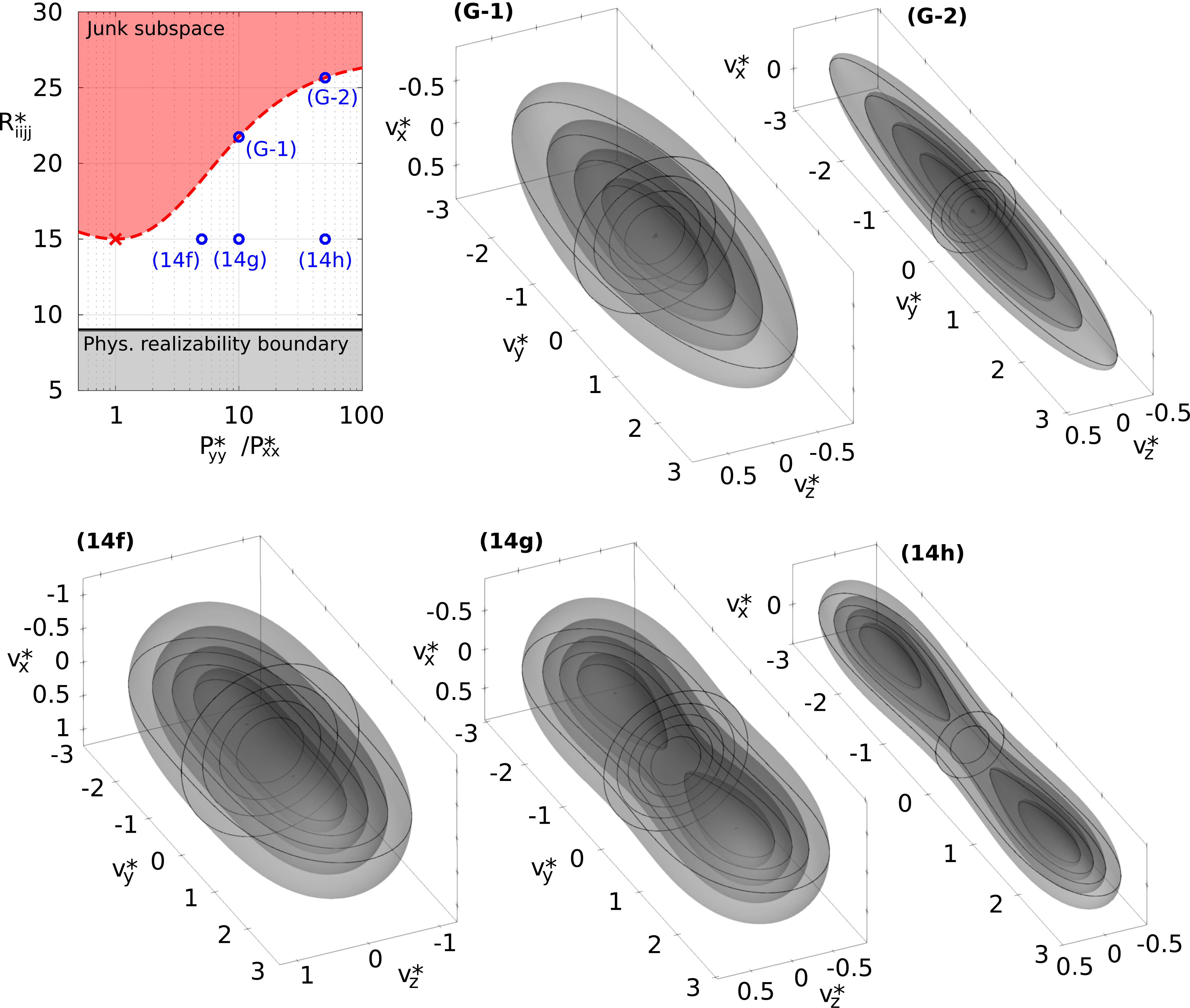}
\caption{Top-Left: Junk subspace, physical realizability boundary and probed points in moment space.
         Other boxes: contours of the 14-moment maximum-entropy VDF in dimensionless velocity space, for two Gaussians, 
         (G-1) and (G-2), and for cases (14f), (14g) and (14h). 
         $R_{iijj}^\star$ and $Q_{ijk}^\star$ are taken at equilibrium, while the pressure anisotropy is increased from
         Left to Right.}
\label{fig:14mom-f-Press-anis}
\end{figure}

The choice of $R_{iijj}^\star$ needs a word of caution.
As discussed, in the presence of pressure anisotropy, the Junk subspace deforms, and the value of $R_{iijj}^\star$ at its lower end corresponds
to that of a Gaussian distribution. 
We always have $R_{iijj}^{\star\, \mathcal{G}} \ge R_{iijj}^{\star \, \mathcal{M}}$.
This is also shown in Fig.~\ref{fig:14mom-f-Press-anis}-Top-Left.
In the present section, we thus consider the following states:
\begin{itemize}
  \item two Gaussian states, (G-1) and (G-2), are selected at the lower end of the Junk subspace;
  \item other three states, (14f), (14g) and (14h) are instead selected with a fourth-order moment fixed to the Maxwellian value, $R_{iijj}^\star = 15$.
\end{itemize}

\noindent These states are:
\begin{description}
  \item[\rm(G-1)] $P_{xx}^\star = P_{zz}^\star = P_{yy}^\star/10 \ \ , \ \ \  R_{iijj}^\star = R_{iijj}^{\star \, \mathcal{G}} \ \ , \ \ \ Q^\star_{ijk} = 0\ $; 
  \item[\rm(G-2)] $P_{xx}^\star = P_{zz}^\star = P_{yy}^\star/50 \ \ , \ \ \  R_{iijj}^\star = R_{iijj}^{\star \, \mathcal{G}} \ \ , \ \ \ Q^\star_{ijk} = 0\ $; 
  \item[\rm(14f)] $P_{xx}^\star = P_{zz}^\star = P_{yy}^\star/5 \ \ , \ \ \  R_{iijj}^\star = 15 \ \ , \ \ \ Q^\star_{ijk} = 0\ $; 
  \item[\rm(14g)] $P_{xx}^\star = P_{zz}^\star = P_{yy}^\star/10 \ \ , \ \ \  R_{iijj}^\star = 15 \ \ , \ \ \ Q^\star_{ijk} = 0\ $;
  \item[\rm(14h)] $P_{xx}^\star = P_{zz}^\star = P_{yy}^\star/50 \ \ , \ \ \  R_{iijj}^\star = 15 \ \ , \ \ \ Q^\star_{ijk} = 0\ $.
\end{description}

\noindent These cases present a degree of anisotropy of 5, 10 and 50.
The resulting 14-moment ME distributions are shown in Fig.~\ref{fig:14mom-f-Press-anis}.
As expected, for cases (G-1) and (G-2), the 14-moment VDF recovers the anisotropic Gaussian, whose iso-surfaces are ellipsoids 
(Fig.~\ref{fig:14mom-f-Press-anis}-Top-Centre and Top-Right).
On the other hand, prescribing a sub-Gaussian value for $R_{iijj}^\star$, results in the distribution splitting in two parts.
These peaks can be made asymmetric by introducing a non-zero heat flux, as done in Section~\ref{sec:14-mom-Pij-Qijj}.

\begin{figure}[h]
\centering
\includegraphics[width=0.6\textwidth]{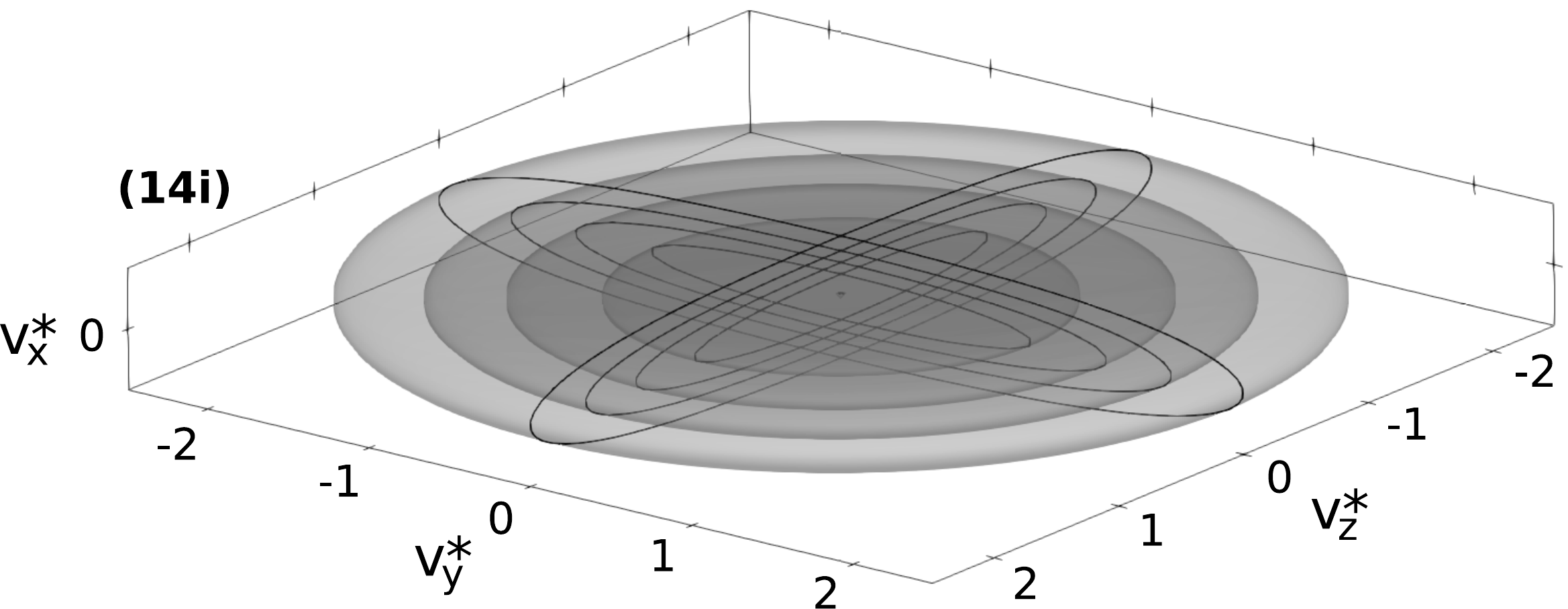}
\caption{Contours of the 14-moment maximum-entropy VDF in dimensionless velocity space, for case (14i).
         $R_{iijj}^\star$ and $Q_{ijj}^\star$ are taken at equilibrium. The VDF is ``cold'' along $x$, while the $y$ and $z$
         directions are associated with a higher temperature (pressure).}
\label{fig:14mom-f-Press-anis-PP}
\end{figure}

Figure~\ref{fig:14mom-f-Press-anis} shows the evolution of the ME VDF as the pressure is increased along a single axis.
Instead, if the pressure along \textit{two} different axes is larger than the third one, the distribution takes the shape of a flattened
spheroid, as shown in Fig.~\ref{fig:14mom-f-Press-anis-PP}, for the following condition,
\begin{description}
  \item[\rm(14i)] $P_{yy}^\star = P_{zz}^\star = 50 \, P_{xx}^\star \ \ , \ \ \  Q^\star_{ijj} = 0\ \ , \ \ \ R_{iijj}^\star = 15 \ $.
\end{description}

\noindent Finally, the case where the three pressures are individually different can be seen to be a gradual 
transition between cases (14h) and (14i).

\subsection{Combined effect of $P_{ij}$ and $Q_{ijj}$}\label{sec:14-mom-Pij-Qijj}

The anisotropic ME distributions of Fig.~\ref{fig:14mom-f-Press-anis} can be made asymmetric by introducing a heat flux.
Let us consider a gas state characterized by an equilibrium fourth-order moment, $R_{iijj}^\star = 15$, 
and a degree of anisotropy of 50, with $P_{xx}^\star=P_{zz}^\star=P_{yy}^\star/50$.
This results in the slender VDF of case (14h), shown in Fig.~\ref{fig:14mom-f-Press-anis}.

\subsubsection{Effect of the axial heat flux}\label{sec:cases-14h-num-axial-heat-flux}

\begin{figure}[h]
\centering
\includegraphics[width=\textwidth]{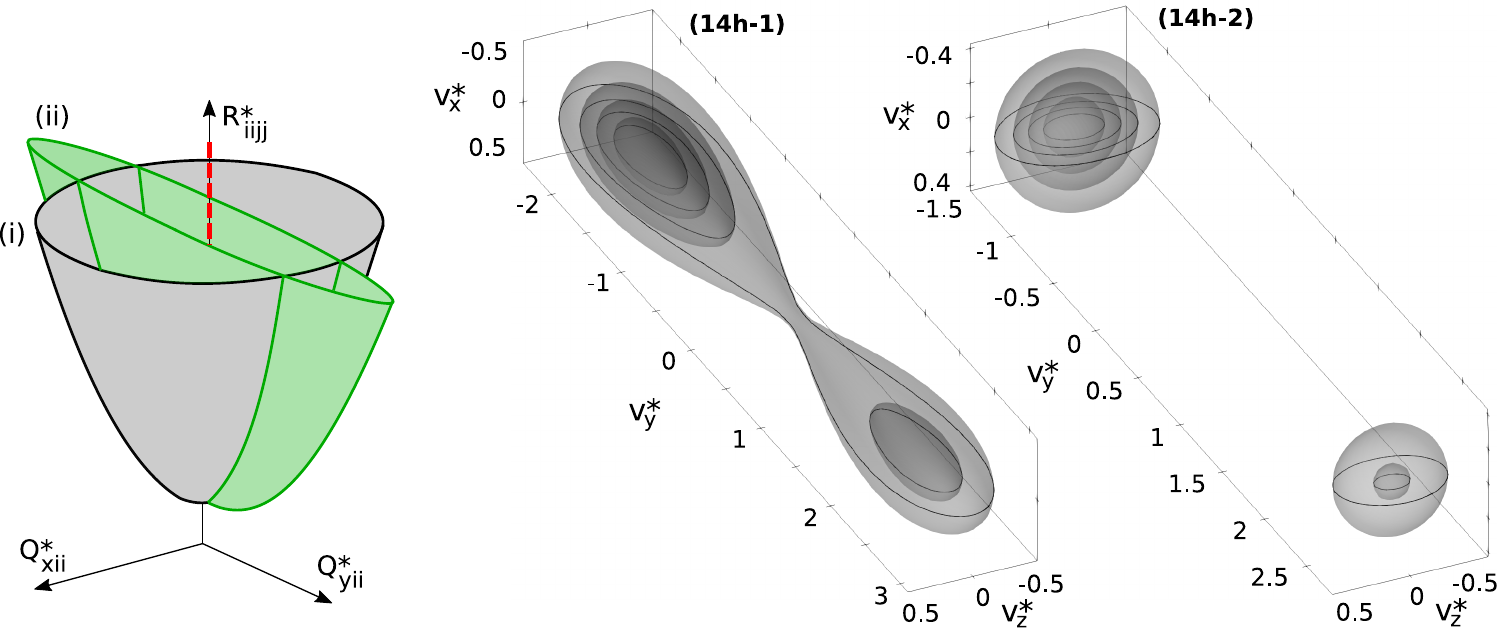}
\caption{Left: physical realizability boundary from Eq.~\eqref{eq:realizability-14mom-Riijj-dimensionless}
         for (i) an isotropic gas state, and for (ii) an anisotropic state with $P_{xx}^\star=P_{zz}^\star=P_{yy}^\star/50$. 
         Centre and Right: effect of a non-zero heat flux on the baseline VDF of case (14h), shown in Fig.~\ref{fig:14mom-f-Press-anis}.}
\label{fig:14mom-f-Press-anis-Qyjj-2maxw}
\end{figure}

In case (14h), the pressure (and thus the energy) is mostly concentrated on the $y$ axis.
Therefore, it can be expected that this direction can also support a larger heat flux, if compared to the other two, colder, directions.
This can be formally demonstrated by analyzing the realizability boundary of Eq.~\eqref{eq:realizability-14mom-Riijj-dimensionless}.
In the presence of anisotropy, the realizability boundary deforms, as shown in Fig.~\ref{fig:14mom-f-Press-anis-Qyjj-2maxw}-Left.
For the said anisotropy, and for $R^\star_{iijj}=15$, Eq.~\eqref{eq:realizability-14mom-Riijj-dimensionless} predicts that 
the maximum realizable heat flux is $Q_{yjj}^{\star\, \mathrm{max}} \approx 4.16$.

In order to study the effect of the heat flux along the slenderness direction, $y$,
we consider the following cases, derived from case (14h):
\begin{description}
  \item[\rm(14h-1)] $P_{xx}^\star=P_{zz}^\star=P_{yy}^\star/50 \ , \ Q_{yjj}^\star = 2   \ , \ Q_{xjj}^\star = Q_{zjj}^\star = 0\ , \ R_{iijj}^\star = 15 \ $;
  \item[\rm(14h-2)] $P_{xx}^\star=P_{zz}^\star=P_{yy}^\star/50 \ , \ Q_{yjj}^\star = 4  \ , \ Q_{xjj}^\star = Q_{zjj}^\star = 0\ , \ R_{iijj}^\star = 15 \ $;
  \item[\rm(14h-3)] $P_{xx}^\star=P_{zz}^\star=P_{yy}^\star/50 \ , \ Q_{yjj}^\star = 4.12 \ , \ Q_{xjj}^\star = Q_{zjj}^\star = 0 \ , \ R_{iijj}^\star = 15 \ $.
\end{description}

\begin{figure}[h]
\centering
\includegraphics[width=0.7\textwidth]{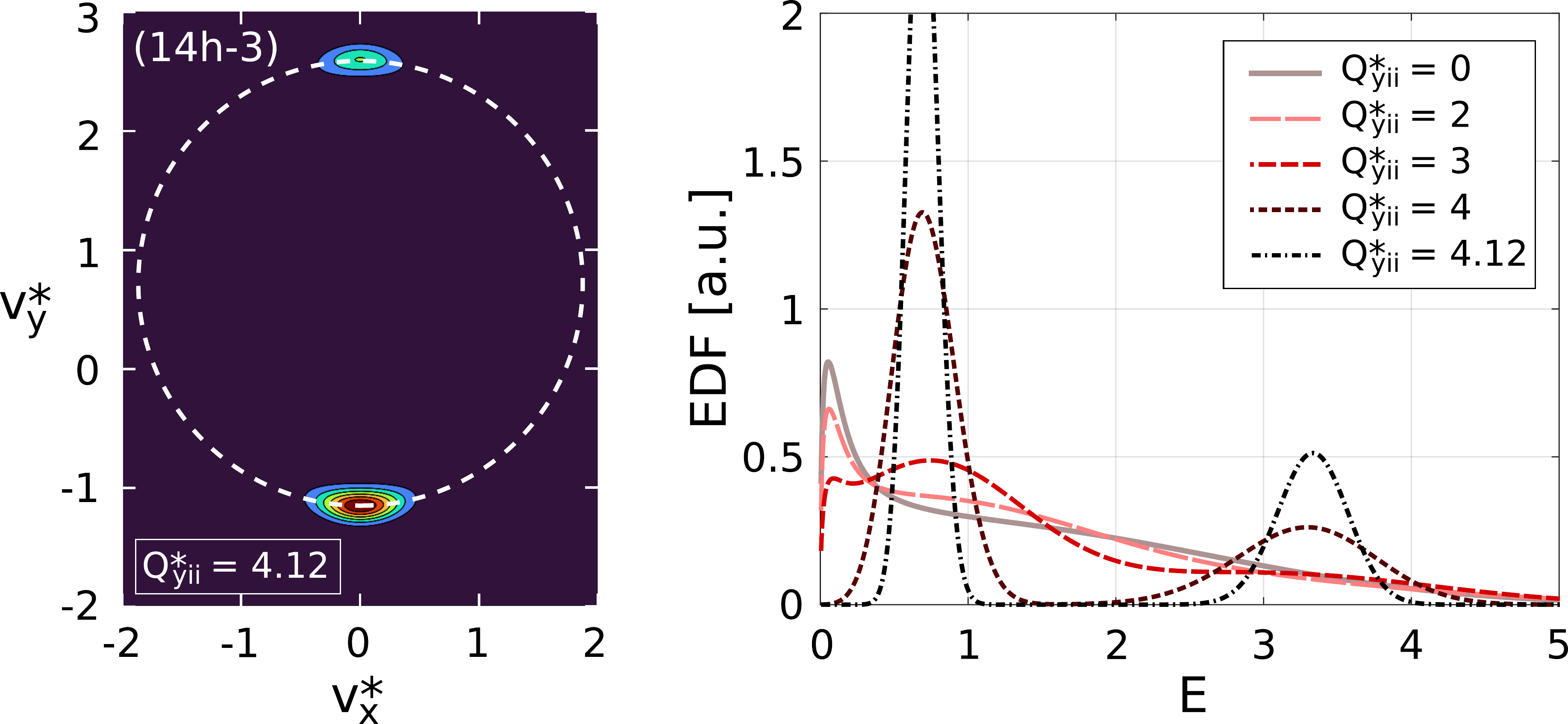}
\caption{Velocity and energy distribution functions for a strongly anisotropic gas state, with $P_{xx}=P_{zz}=P_{yy}/50$, 
         and with $R_{iijj}^\star=15$.
         Left: slice in the $(v_x^\star, v_y^\star)$ plane of the 14-moment ME VDF from case (14h-3).
         The white dashed line represents the sphere from Eq.~\eqref{eq:radius-sphere-velspace-14mom}. 
         Right: ME EDFs for four different heat fluxes.}
\label{fig:14mom-f-Pij-Qyjj-EDFs}
\end{figure}

\noindent The first two cases are shown in Fig.~\ref{fig:14mom-f-Press-anis-Qyjj-2maxw}-Centre and Right,
while the third one is shown in Fig.~\ref{fig:14mom-f-Pij-Qyjj-EDFs}-Left.
As the VDFs split into two separate peaks, the EDFs also assume a bi-modal shape (see Fig.~\ref{fig:14mom-f-Pij-Qyjj-EDFs}-Right). 
Close to the realizability boundary, the two contributions to the EDFs asymptote to sharp distributions, resembling Dirac deltas.
From a physical perspective, the distribution functions of Figures~\ref{fig:14mom-f-Press-anis-Qyjj-2maxw}
and \ref{fig:14mom-f-Pij-Qyjj-EDFs} can represent situations of a fast-moving particle beam travelling
through a denser background.

Notice that, from a graphical inspection, the two separate peaks of the VDFs appear to have roughly the same temperature. 
This is a recurrent feature of maximum-entropy methods, and can be easily appreciated when considering one-dimensional formulations in
the velocity, see for instance \cite{laplante2016comparison}.
The same feature emerges in the 14-moment model.


\subsubsection{Effect of the transverse heat flux}

\begin{figure}[h]
\centering
\includegraphics[width=0.9\textwidth]{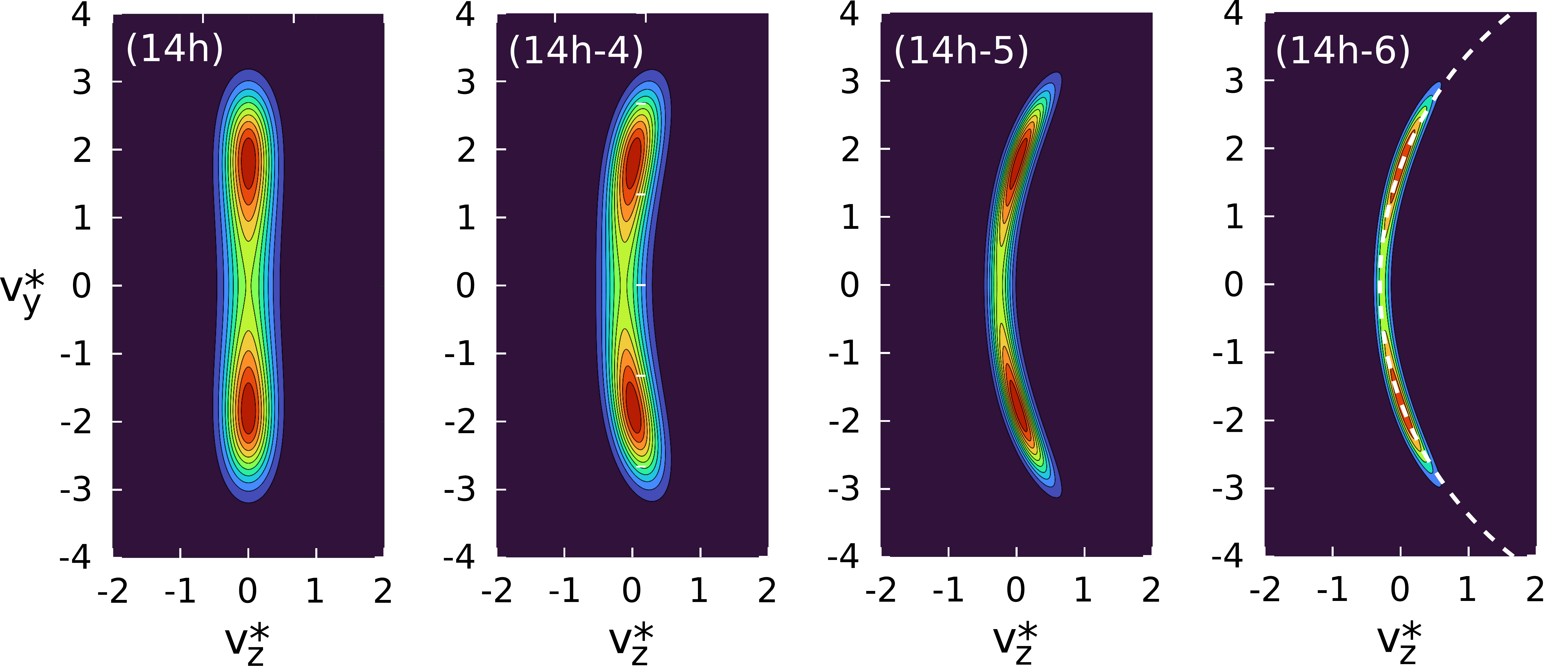}
\caption{Effect of a transverse heat flux on the anisotropic VDF of case (14h); 
         slices on the $(v^\star_y, v^\star_z)$ plane, passing by $v_x^\star=0$. 
         The VDF of case (14h) is symmetric, as no heat flux is present. Cases (14h-4), (14h-5) and (14h-6) present
         an increasing heat flux, and approach the physical realizability boundary.}
\label{fig:14mom-f-Pij-Qxjj-transverse}
\end{figure}

Let us consider again case (14h) as a baseline.
As an effect of the anisotropy, the maximum heat flux along the transverse directions, $x$ and $z$,
reduces to $Q_{xjj}^{\star \, \mathrm{max}}=Q_{zjj}^{\star \, \mathrm{max}}\approx 0.5883$.
The effect of a non-zero transverse heat flux is shown in Fig.~\ref{fig:14mom-f-Pij-Qxjj-transverse}, where,
beside case (14h), we consider the following states, the latter being the closest to the realizability boundary:
\begin{description}
  \item[\rm(14h-4)] $P_{xx}^\star=P_{zz}^\star=P_{yy}^\star/50 \ , \ Q_{xjj}^\star = 2.5 \ , \ Q_{yjj}^\star = Q_{zjj}^\star = 0 \ , \ R_{iijj}^\star = 15 \ $;
  \item[\rm(14h-5)] $P_{xx}^\star=P_{zz}^\star=P_{yy}^\star/50 \ , \ Q_{xjj}^\star = 5   \ , \ Q_{yjj}^\star = Q_{zjj}^\star = 0 \ , \ R_{iijj}^\star = 15 \ $;
  \item[\rm(14h-6)] $P_{xx}^\star=P_{zz}^\star=P_{yy}^\star/50 \ , \ Q_{xjj}^\star = 5.5 \ , \ Q_{yjj}^\star = Q_{zjj}^\star = 0 \ , \ R_{iijj}^\star = 15 \ $.
\end{description}

\FloatBarrier
\subsection{Combined effect of $P_{ij}$, $Q_{ijj}$ and $R_{iijj}$}

The effect of a sub-Maxwellian kurtosis, with fourth-order moment $R_{iijj}^\star < 15$, 
is seen to condense an otherwise equilibrium VDF to a spherical shell (Figures~\ref{fig:14mom-R-Q-VDFs} and \ref{fig:14mom-f-various-R}).
If $R_{iijj}^\star$ is sufficiently small, a hole is formed at the centre of the VDF, as in case (14a).
Combining this with the presence of pressure anisotropy, one can obtain the following shapes.

\subsubsection{Effect on a flat VDF}

\begin{figure}[h]
\centering
\includegraphics[width=0.6\textwidth]{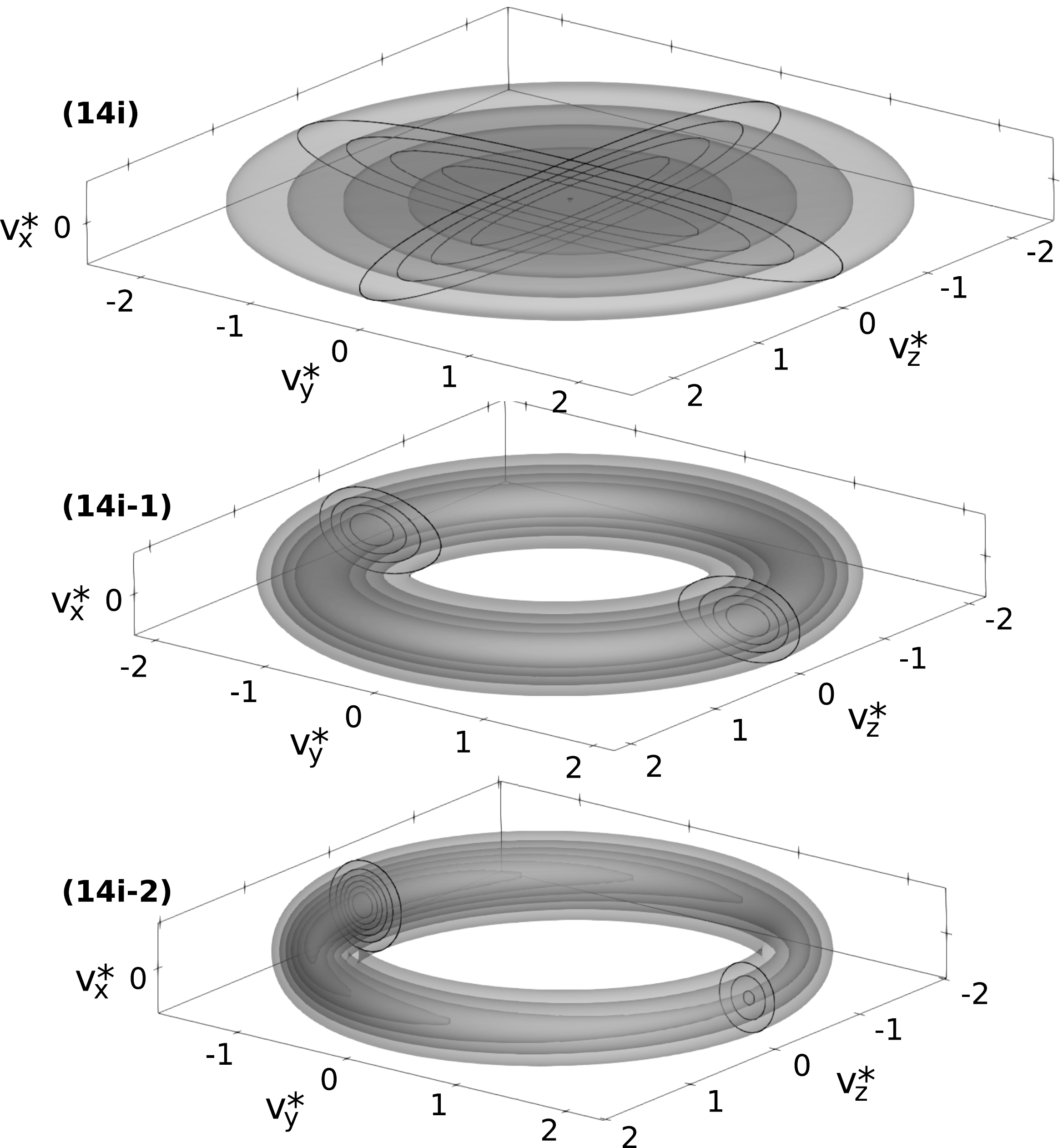}
\caption{Modification of the disk-like anisotropic distribution (14i) 
         under the effect of a sub-Maxwellian kurtosis, $R_{iijj}^\star = 10$ (case 14i-1),
         combined with a non-zero heat flux, $Q_{yii}^\star = 1$, (case 14i-2).}
\label{fig:14mom-f-Pyy-flat-R10-torus}
\end{figure}

A disk-like VDF, such as that of case (14i), under the effect of a sufficiently small value of $R_{iijj}^\star$, transforms into a torus.
The further presence of a heat flux introduces anisotropy.
Figure~\ref{fig:14mom-f-Pyy-flat-R10-torus} shows the case (14i), and the two additional cases
\begin{description}
  \item[\rm(14i-1)] $P_{yy}^\star = P_{zz}^\star = 50 \, P_{xx}^\star \ , \  Q^\star_{ijj} = 0\ , \ R_{iijj}^\star = 10 \ $.
  \item[\rm(14i-2)] $P_{yy}^\star = P_{zz}^\star = 50 \, P_{xx}^\star \ , \  Q^\star_{xjj} = 1 \ , \ Q^\star_{yjj} = Q^\star_{zjj} = 0 \ , \ R_{iijj}^\star = 10 \ $.
\end{description}

\begin{figure}[h]
\centering
\includegraphics[width=1.0\textwidth]{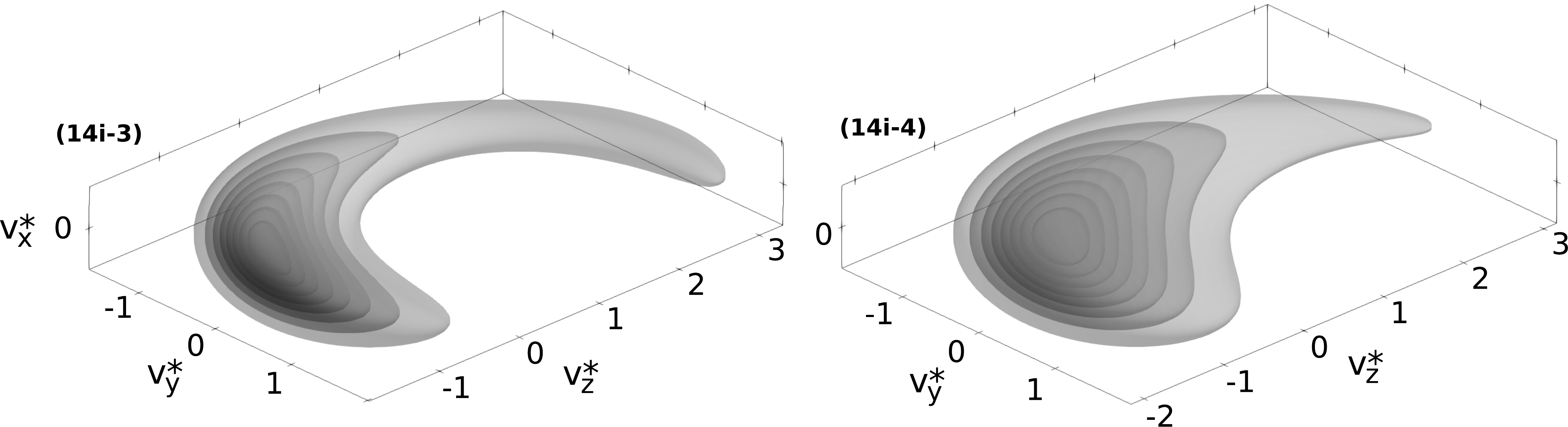}
\caption{14-moment VDF for the cases (14i-3) and (14i-4), with $P_{yy}^\star > P_{zz}^\star > P_{xx}^\star$, 
         super-Maxwellian kurtosis and a non-zero heat flux along both $y$ and $z$.}
\label{fig:14mom-f-Pyy-flat-R21-cotoletta}
\end{figure}

\noindent Further asymmetry can be added to this problem by introducing a non-zero heat flux along $y$, and by
changing the ratio between $P_{yy}$ and $P_{zz}$.
By a proper modification of these parameters, one can obtain toroidal asymmetric distributions 
analogous to those observed in \cite{boccelli202014}.
Yet other shapes can be obtained by employing a super-Maxwellian kurtosis, with $R_{iijj}^\star > 15$,
combined with a heat flux and different degrees of anisotropy.
As an example, Figure~\ref{fig:14mom-f-Pyy-flat-R21-cotoletta} shows two cases where $P_{yy}^\star > P_{zz}^\star > P_{xx}^\star$, 
\begin{description}
  \item[\rm(14i-3)] $P_{yy}^\star =2\, P_{zz}^\star = 50\, P_{xx}^\star \ , \ Q_{xjj}^\star=0 \ , \  Q_{yjj}^\star = 3 \ , \ Q_{zjj}^\star = 1.5 \ , \ R_{iijj}^\star = 18 \ $;
  \item[\rm(14i-4)] $P_{yy}^\star =2\, P_{zz}^\star = 50\, P_{xx}^\star \ , \ Q_{xjj}^\star = 0 \ , \ Q_{yjj}^\star = 3 \ , \ Q_{zjj}^\star = 1.5 \ , \ R_{iijj}^\star = 21 \ $.
\end{description}

\subsubsection{Effect on an elongated VDF}

We analyze here the evolution of an elongated VDF with $P_{yy}>P_{zz}=P_{xx}$, in the presence of both $R_{iijj}^\star < 15$ and of a heat flux,
transverse with respect to the elongation direction.
We consider here the cases
\begin{description}
  \item[\rm(14j-1)] $P_{xx}^\star=P_{zz}^\star=P_{yy}^\star/2 \ , \ Q_{xjj}^\star = 0    \ , \ Q_{yjj}^\star = Q_{zjj}^\star = 0 \ , \ R_{iijj}^\star = 10 \ $;
  \item[\rm(14j-2)] $P_{xx}^\star=P_{zz}^\star=P_{yy}^\star/2 \ , \ Q_{xjj}^\star = 0.5  \ , \ Q_{yjj}^\star = Q_{zjj}^\star = 0 \ , \ R_{iijj}^\star = 10 \ $;
  \item[\rm(14j-3)] $P_{xx}^\star=P_{zz}^\star=P_{yy}^\star/2 \ , \ Q_{xjj}^\star = 0.75 \ , \ Q_{yjj}^\star = Q_{zjj}^\star = 0 \ , \ R_{iijj}^\star = 10 \ $.
\end{description}

\begin{figure}[h]
\centering
\includegraphics[width=0.9\textwidth]{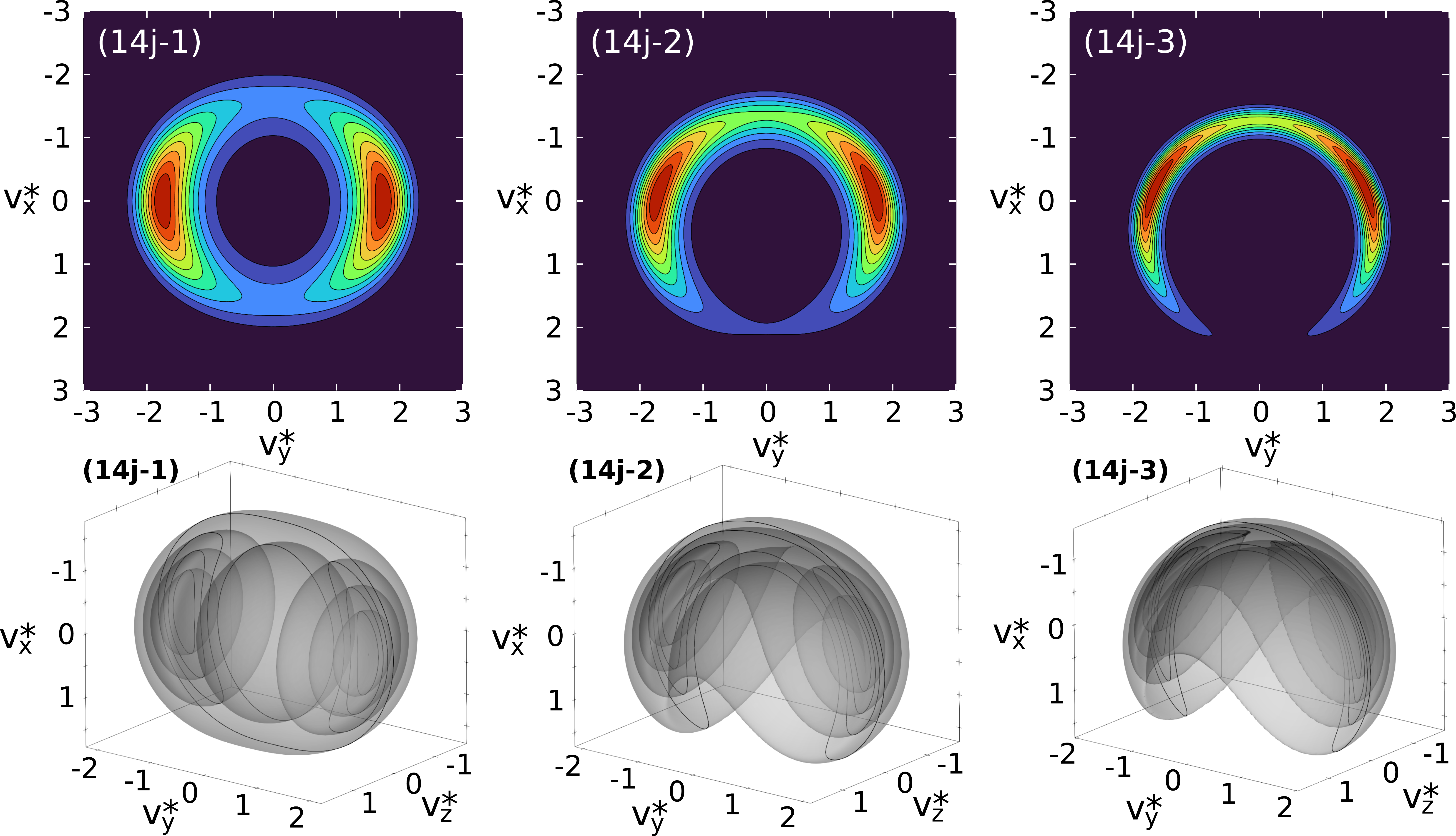}
\caption{Effect of $R_{iijj}^\star$ and of a transverse heat flux, $Q_{xii}^\star$, on an anisotropic VDF.
         Top: slices of the 14-moment ME VDF on the $(v_x^\star, v_y^\star)$ plane. Bottom: contours of the same VDFs 
         in three-dimensional velocity space.}
\label{fig:14mom-f-Pyy-R10-Qx}
\end{figure}

\noindent These cases present a degree of anisotropy of 2, have a sub-Maxwellian kurtosis, $R_{iijj}^\star = 10$, and present a progressively
increasing heat flux.
The resulting VDFs are shown in Fig.~\ref{fig:14mom-f-Pyy-R10-Qx}.

\subsection{Approaching the Junk subspace}\label{sec:14-mom-Junk-subspace-VDFs}

As discussed in Section~\ref{sec:phys-real-boundary-and-junk}, 
the entropy maximization problem has a unique solution everywhere in the physically realizable moment space,
except in correspondence with the Junk subspace. 
As the gas state approaches the Junk subspace, the partial differential equations associated with the maximum-entropy 
method encounter a singularity.
In particular, some closing moments appearing in the spacial fluxes assume an infinite value, 
and the eigenvalues of the flux Jacobian consequently also diverge.

It is interesting to analyze how the distribution function appears in such situations.
As discussed in \cite{mcdonald2016approximate} for one-dimensional distributions, 
near the Junk subspace the ME VDF is composed of a quasi-Maxwellian bulk and by a small-amplitude bump located along one of the tails.
As the Junk subspace is approached from one side, this bump:
\begin{itemize}
  \item Gets farther from the bulk, and quickly reaches large velocities;
  \item Decreases in amplitude, and one might need to consider a logarithmic scaling of the VDF in order to visualize it.
\end{itemize}

\begin{figure}[h]
\centering
\includegraphics[width=\textwidth]{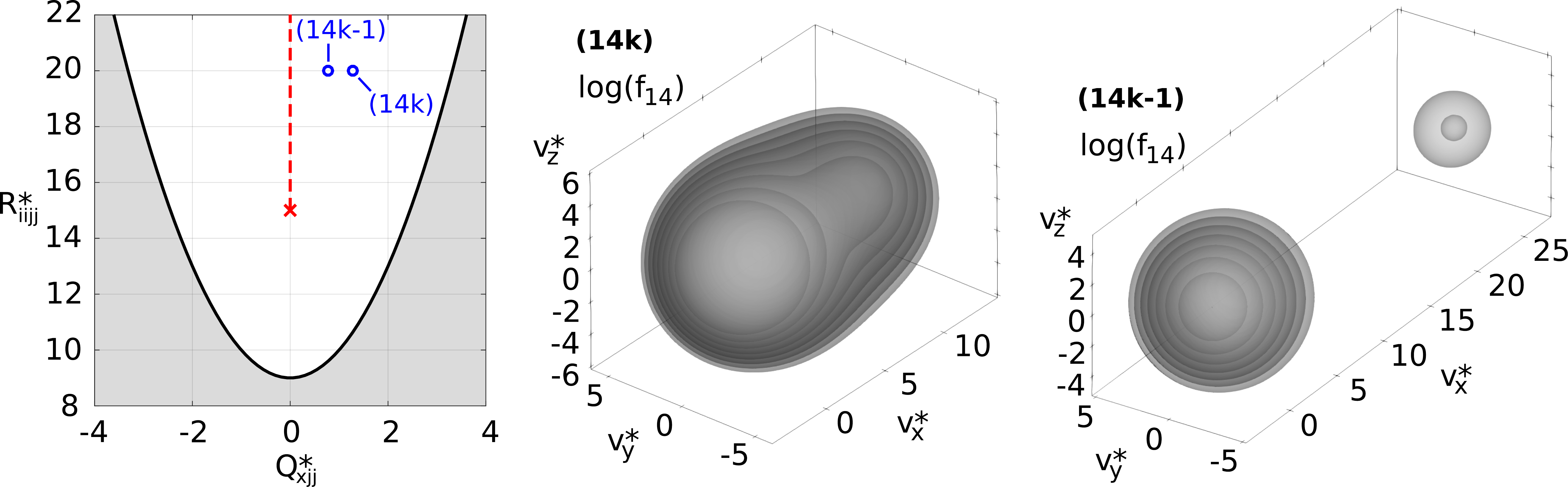}
\caption{Contours of $\log(f_{14})$, for cases (14k) and (14k-1). As the Junk subspace is approached from one side, the ME VDF appears as a quasi-Maxwellian
bulk, with a dim high-velocity bump on one side. A logarithmic scaling of the VDF is necessary for visualizing the high-velocity bump.}
\label{fig:14mom-f-Junk}
\end{figure}

These observations also apply to the three-dimensional 14-moment distribution, as shown in Figure~\ref{fig:14mom-f-Junk}, for the 
states:
\begin{description}
  \item[\rm(14k)]  \hphantom{-1}$P_{ij}^\star= \delta_{ij}\ , \ Q_{xjj}^\star = 1.28 \ , \ Q_{yjj}^\star = Q_{zjj}^\star = 0 \ , \ R_{iijj}^\star = 20 \ $;
  \item[\rm(14k-1)]  $P_{ij}^\star= \delta_{ij}\ , \ Q_{xjj}^\star = 0.774 \ , \ Q_{yjj}^\star = Q_{zjj}^\star = 0 \ , \ R_{iijj}^\star = 20 \ $;
\end{description}

\noindent In these states, we consider pressure isotropy for simplicity, and approach the Junk subspace from one side.
In case (14k), the ME VDF appears as a Maxwellian bulk, deformed along one side.
At lower values of the heat flux the deformation increases, eventually evolving into a separate bump (case 14k-1).
Notice that the VDFs shown in the plots are scaled logarithmically.
Indeed, a linear scaling would only show the quasi-Maxwellian bulk, and no trace of the side bump would be visible.

\section{The 21-moment maximum-entropy moment method}\label{sec:21mom-maxent}

The 21-moment ME VDF reads
\begin{equation}\label{eq:21mom-VDF-expdefinition}
  f^{\mathrm{ME}}_{21} = \exp\left( \alpha_0 + \alpha_i v_i + \alpha_{ij} v_i v_j + \alpha_{ijk} v_i v_j v_k + \alpha_{4} v^4 \right) \, .
\end{equation}

\noindent This is still a fourth-order distribution in velocity.
The respective vector of moments of interest, $\bm{U}_{21}$, is
\begin{equation}
  \bm{U}_{21} 
  = 
    \begin{pmatrix} 
       \rho  \\ 
       \rho u_i  \\ 
       \rho u_i u_j + P_{ij} \\
       \rho u_i u_j u_k + u_i P_{jk} + u_j P_{ki} + u_k P_{ij} + Q_{ijk} \\
       \rho u_i u_i u_j u_j + 2 u_i u_i P_{jj} + 4 u_i u_j P_{ij} + 4 u_i Q_{ijj} + R_{iijj} 
    \end{pmatrix} \, .
\end{equation}

The vector of primitive variables of interest, in dimensionless form, and considering a reference frame centred on the bulk velocity, is
\begin{equation}
  \bm{W}_{21}^\star = \left(\rho^\star = 1, u_i^\star = 0, P_{ij}^\star, Q_{ijk}^\star, R_{iijj}^\star \right)\, .
\end{equation}

Denoting by $\mathbb{D}_{14}$ the set of all possible 14-moment VDFs, we have that $\mathbb{D}_{14} \subset \mathbb{D}_{21}$.
The 21-moment method grants the additional freedom to modify the components of the
heat flux tensor, $Q_{ijk}$, individually, provided that the physical realizability condition is still met.
When these components are equal to the value that they would take in the 14-moment maximum-entropy model, 
then the 21 and the 14-moment VDFs coincide.
Otherwise, different non-equilibrium shapes emerge, as investigated in the following.


\subsection{VDFs with $C_3$ symmetry and a mirror plane}\label{sec:21mom-C3-symmetry}

\begin{figure}[h]%
\centering
\includegraphics[width=1.0\textwidth]{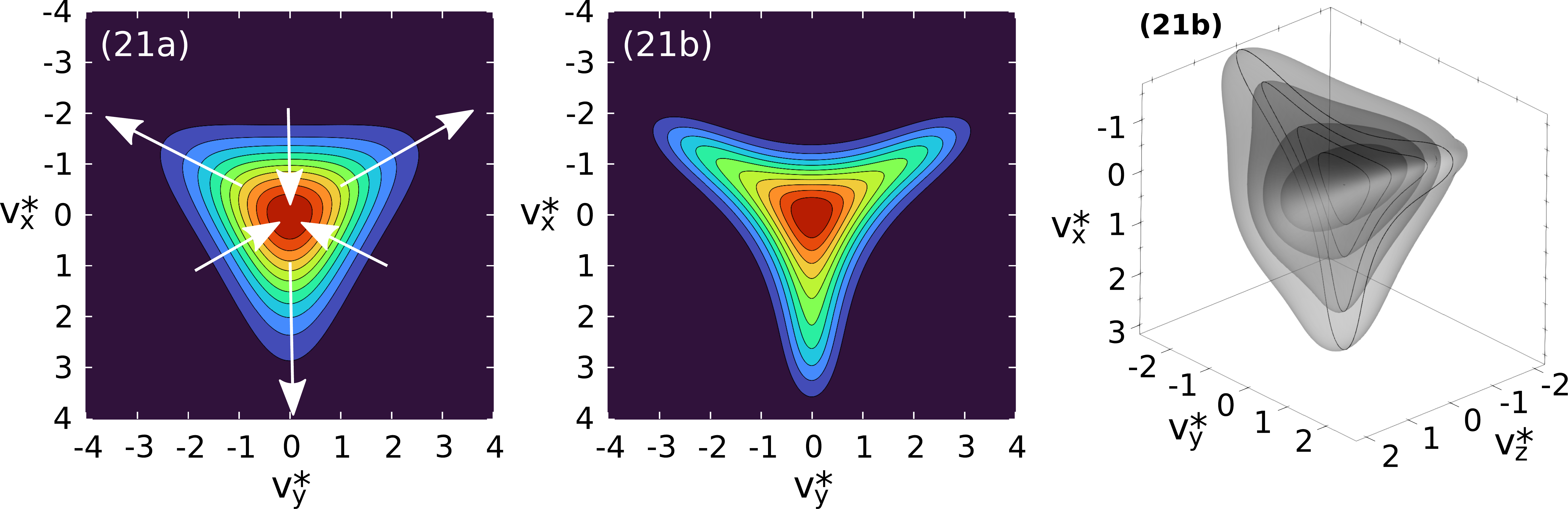}
\caption{21-moment ME VDFs for cases (21a) and (21b). The heat flux tensor components are balanced, and give a zero net heat flux vector. The Left and Centre panels show a slice of the VDF, for $v_z^\star=0$. The Right panel shows the same VDF as the centre panel, represented as three-dimensional contours. 
White arrows are superimposed, to illustrate the effect of the considered heat flux terms, on an otherwise Maxwellian VDF.}
\label{fig:VDF21-effect-Q-triangle}
\end{figure}

The simplest way to build a 21-moment VDF that cannot be reproduced by the 14-moment model consists of manufacturing a heat flux tensor
whose traces are zero ($Q_{ijj} = 0$), yet its individual components are not.
Consider for instance the following two gas states,
\begin{description}
  \item[\rm(21a)] $P_{ij}^\star = \delta_{ij} \ \ , \ \ \ Q_{xxx}^\star = - Q_{xyy}^\star = 0.5 \ \ , \ \ \ R_{iijj}^\star = 15 \ $;
  \item[\rm(21b)] $P_{ij}^\star = \delta_{ij} \ \ , \ \ \ Q_{xxx}^\star = - Q_{xyy}^\star = 0.8 \ \ , \ \ \ R_{iijj}^\star = 15 \ $;
\end{description}

\noindent all other entries in the heat flux tensor being zero.
For these states, the $x$-component of the heat flux \textit{vector} 
is $Q_{xjj} = Q_{xxx} + Q_{xyy} + Q_{xzz} = 0$, and the 14-moment model would thus predict a 
fully symmetric state.
These cases are shown in Fig.~\ref{fig:VDF21-effect-Q-triangle}.
A 21-moment VDF of this shape was previously observed in the scenario of magnetic reconnection in plasmas, in \cite{ng2018using}.

\begin{figure}[h]%
\centering
\includegraphics[width=0.9\textwidth]{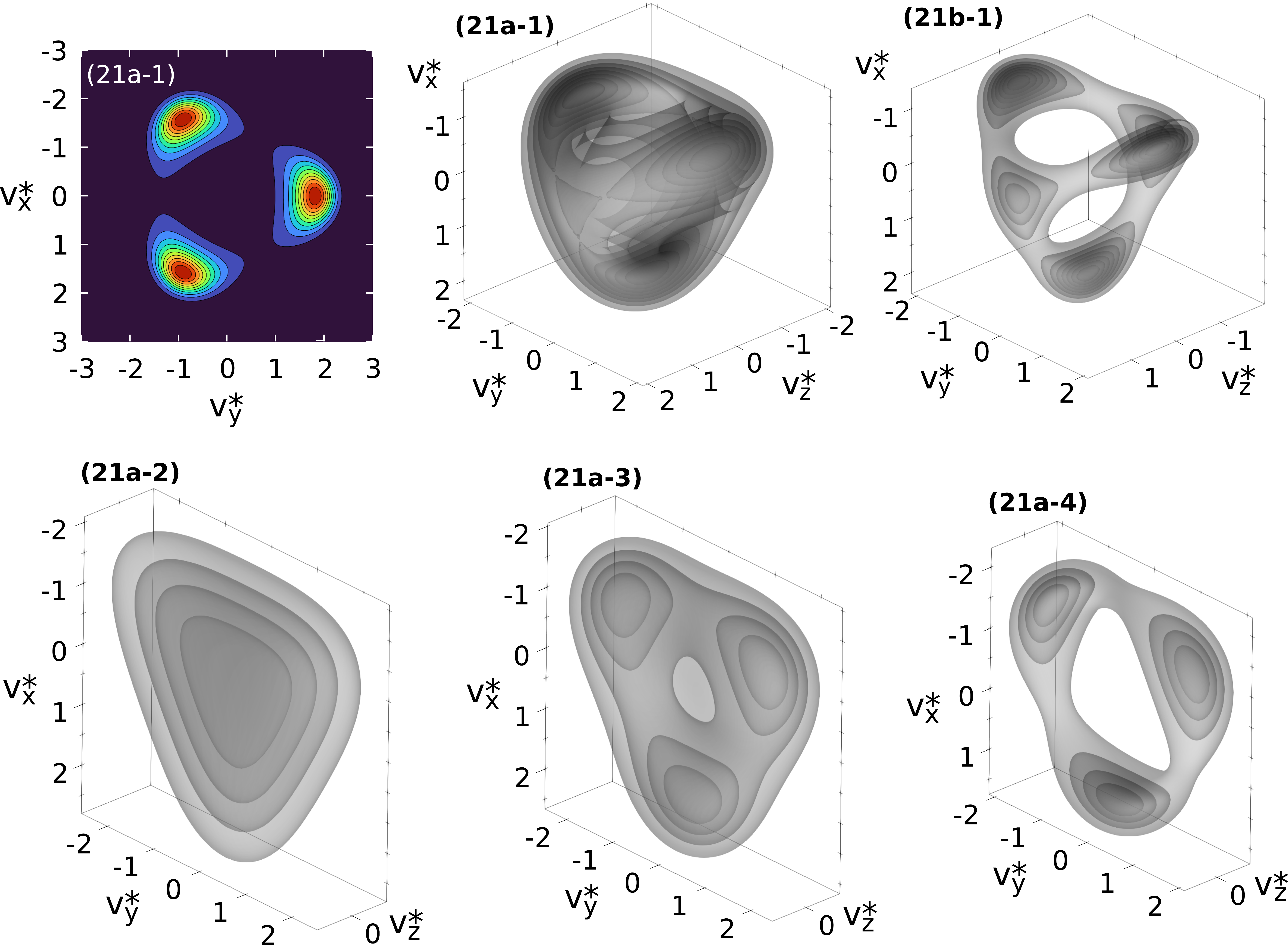}
\caption{21-moment VDFs with $C_3$ rotational symmetry along the $v_z^\star$ axis, obtained as modifications of the baseline cases (21a) and (21b). 
         Notice that the Top-Left and Top-Centre panels show the same VDF, the former representing a slice on the $v_z^\star=0$ plane.}
\label{fig:VDF21-effect-Q-triangle-R10-and-flat}
\end{figure}

These VDFs are characterized by a rotational symmetry of $120^\circ$ ($C_3$ symmetry), around the $v_z^\star$ axis.
Also, they possess a symmetry plane.
Other distributions characterized by the same symmetry can be obtained by lowering the value of $R_{iijj}^\star$ and by introducing pressure anisotropy.
Starting from the baseline cases (21a) and (21b), if $R_{iijj}^\star$ is decreased sufficiently, the VDF condenses in three branches, 
and five local maxima emerge.
This is the case for the following states:

\begin{description}
  \item[\rm(21a-1)] $P_{ij}^\star = \delta_{ij} \ \ , \ \ \ Q_{xxx}^\star = - Q_{xyy}^\star = 0.5 \ \ , \ \ \ R_{iijj}^\star = 10 \ $;
  \item[\rm(21b-1)] $P_{ij}^\star = \delta_{ij} \ \ , \ \ \ Q_{xxx}^\star = - Q_{xyy}^\star = 0.8 \ \ , \ \ \ R_{iijj}^\star = 10 \ $;
\end{description}

\noindent where all the other heat flux components are zero. 
The respective VDFs are shown in Fig.~\ref{fig:VDF21-effect-Q-triangle-R10-and-flat}-Top.
Two of these maxima can be suppressed by reducing the pressure along the $v_z^\star$ axis, with respect to the other two directions,
as for the following cases:

\begin{description}
  \item[\rm(21a-2)] $P_{xx}^\star=P_{yy}^\star=10\, P_{zz}^\star \ \ , \ \ \ Q_{xxx}^\star = - Q_{xyy}^\star = 0.5 \ \ , \ \ \ R_{iijj}^\star = 15 \ $;
  \item[\rm(21a-3)] $P_{xx}^\star=P_{yy}^\star=10\, P_{zz}^\star \ \ , \ \ \ Q_{xxx}^\star = - Q_{xyy}^\star = 0.5 \ \ , \ \ \ R_{iijj}^\star = 12 \ $;
  \item[\rm(21a-4)] $P_{xx}^\star=P_{yy}^\star=10\, P_{zz}^\star \ \ , \ \ \ Q_{xxx}^\star = - Q_{xyy}^\star = 0.5 \ \ , \ \ \ R_{iijj}^\star = 10 \ $;
\end{description}

\noindent where all the other heat flux components are zero. 
These cases are shown in Fig.~\ref{fig:VDF21-effect-Q-triangle-R10-and-flat}-Bottom.

\subsection{VDFs with tetrahedral symmetry}\label{sec:21-mom-tetrahedron}

A tetrahedron-shaped VDF can be obtained by repeating the considerations of Section~\ref{sec:21mom-C3-symmetry} on all the three axis.
We manufacture the heat flux tensor such that the heat flux \textit{vector} along all three directions, $Q_{ijj}^\star$, is zero,
but we individually select non-zero components,
\begin{equation}\label{eq:casefullsymQ21mom}
  \begin{cases}
    Q_{ijk}^\star = 0.5  \ \ \mathrm{if} \ \ i = j = k \, ,\\
    Q_{ijk}^\star = -0.25 \ \ \mathrm{if} \ \ i \neq j = k \, , \\
    Q_{xyz}^\star = 0 \, .
  \end{cases}
\end{equation}

\noindent In other words, along the $x$ direction, we have $Q_{xxx} = 0.5$, $Q_{xyy} = -0.25$, $Q_{xzz} = -0.25$, and thus $Q_{xii} = 0$.
The situation is analogous along $y$ and $z$.
We consider the following cases:
\begin{description}
  \item[\rm(21c)] \hphantom{-1}$P_{ij}^\star = \delta_{ij}$, $Q_{ijk}^\star$ from Eq.~\eqref{eq:casefullsymQ21mom}, $R_{iijj}^\star = 15 \ $;
  \item[\rm(21c-1)] $P_{ij}^\star = \delta_{ij}$, $Q_{ijk}^\star$ from Eq.~\eqref{eq:casefullsymQ21mom}, $R_{iijj}^\star = 12 \ $;
  \item[\rm(21c-2)] $P_{ij}^\star = \delta_{ij}$, $Q_{ijk}^\star$ from Eq.~\eqref{eq:casefullsymQ21mom}, $R_{iijj}^\star = 10 \ $.
\end{description}

\noindent The VDFs associated with these states are shown in Fig.~\ref{sec:21mom-C3-symmetry}.
As $R_{iijj}^\star$ is decreased, from case (21c), the VDF condenses on the edges of the tetrahedron and results in four local maxima.

\begin{figure}[h]%
\centering
\includegraphics[width=0.9\textwidth]{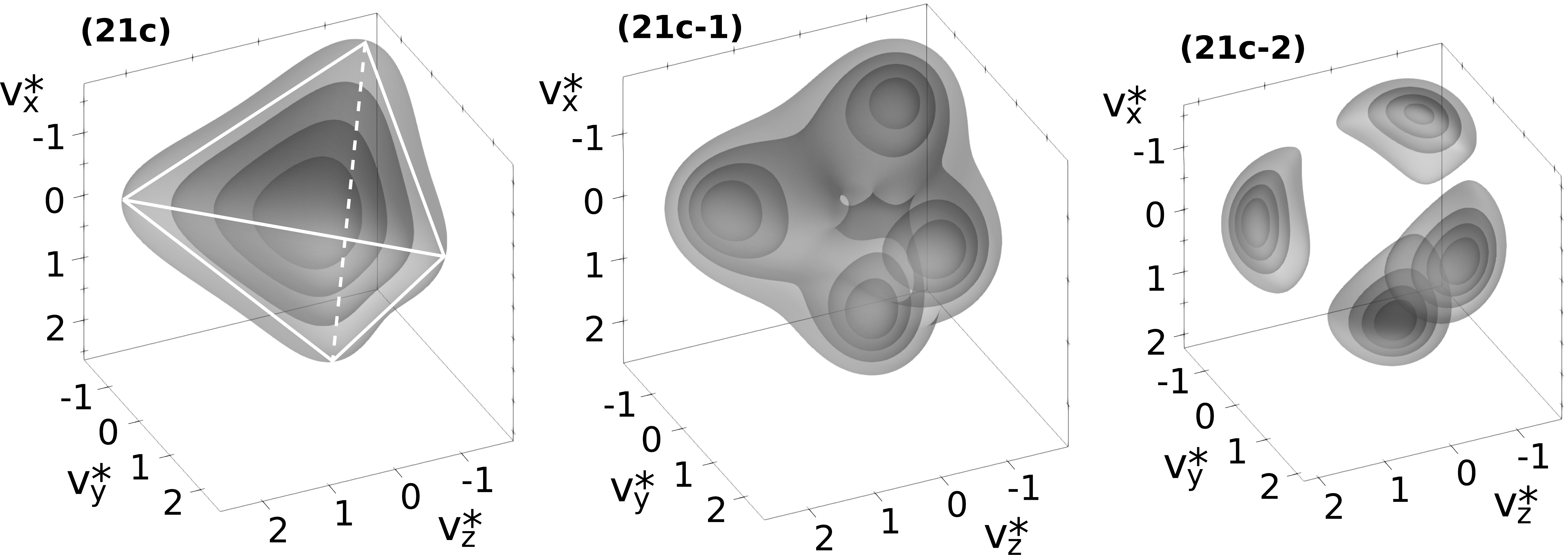}
\caption{21-moment VDFs from cases (21c), (21c-1) and (21c-2). The white lines on case (21c) are a graphical representation of the tetrahedron.}
\label{fig:VDF21-effect-Q-tetrahedron}
\end{figure}


\subsection{Introducing $Q_{xyz}$}\label{sec:21-mom-Qxyx}

\begin{figure}[htpb]%
\centering
\includegraphics[width=0.9\textwidth]{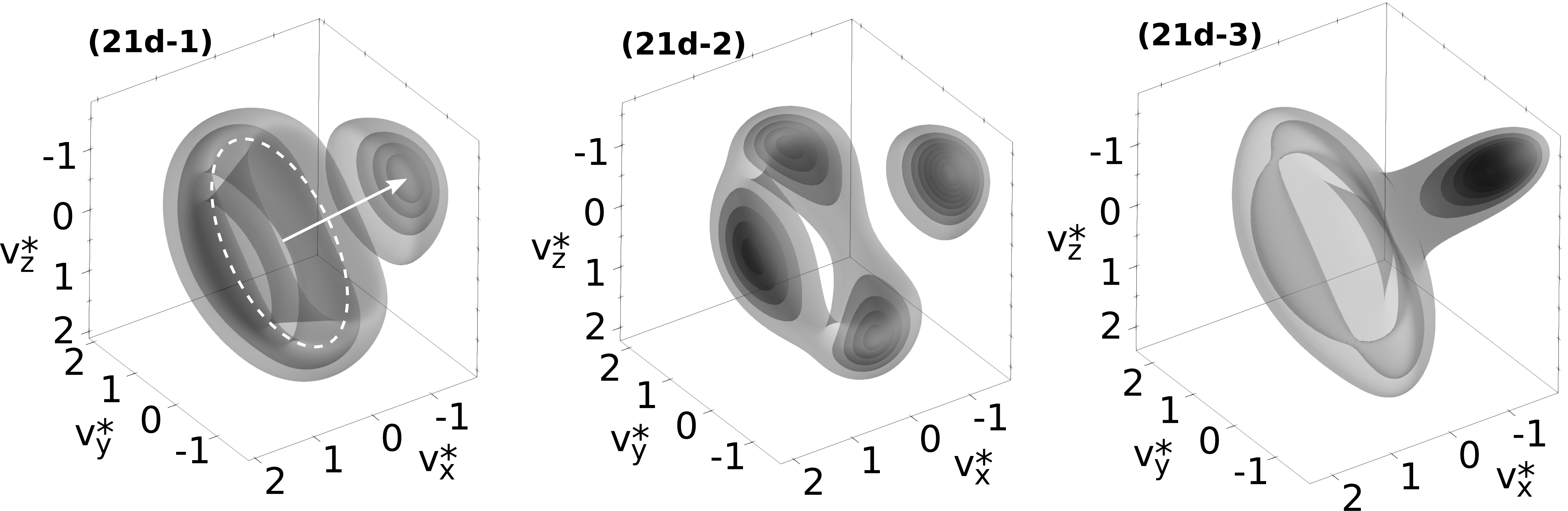}
\caption{21-moment VDFs for cases (21d-1), (21d-2) and (21d-3). 
         The white dashed line and arrow superimposed on case (21d-1) are used to highlight the $C_3$ rotational symmetry.}
\label{fig:VDF21-circle-head}
\end{figure}

Let us introduce a non-zero term, $Q_{xyz}^\star$, on the VDFs of Section~\ref{sec:21-mom-tetrahedron}.
The heat flux tensor is written here by defining two variables, $\beta$ and $\gamma$, such that
\begin{equation}\label{eq:caseQxyz21mom}
  \begin{cases}
    Q_{ijk}^\star = \beta  \ \ \mathrm{if} \ \ i = j = k \, ,\\
    Q_{ijk}^\star = -\beta/2 \ \ \mathrm{if} \ \ i \neq j = k \, , \\
    Q_{xyz}^\star = -\gamma \ \ \mathrm{if} \ \ i \neq j = k \, , \\
  \end{cases}
\end{equation}

\noindent and we consider the following cases: 
\begin{description}
  \item[\rm(21d-1)] $P_{ij}^\star = \delta_{ij}$, $Q_{ijk}^\star$ from Eq.~\eqref{eq:caseQxyz21mom} with $\beta = \gamma = 0.25$, $R_{iijj}^\star = 10 \ $;
  \item[\rm(21d-2)] $P_{ij}^\star = \delta_{ij}$, $Q_{ijk}^\star$ from Eq.~\eqref{eq:caseQxyz21mom} with $\beta = 0.5$, $\gamma = 0.25$, $R_{iijj}^\star = 10 \ $;
  \item[\rm(21d-3)] $P_{ij}^\star = \delta_{ij}$, $Q_{ijk}^\star$ from Eq.~\eqref{eq:caseQxyz21mom} with $\beta = \gamma = 0.5$, $R_{iijj}^\star = 12 \ $.
\end{description}

\noindent The resulting VDFs possess a $C_3$ rotational symmetry along the $(v_x^\star, v_y^\star, v_z^\star)=(1,1,1)$ axis, 
and are shown in Fig.~\ref{fig:VDF21-circle-head}.


\subsection{Other VDF shapes}

\begin{figure}[h]%
\centering
\includegraphics[width=0.9\textwidth]{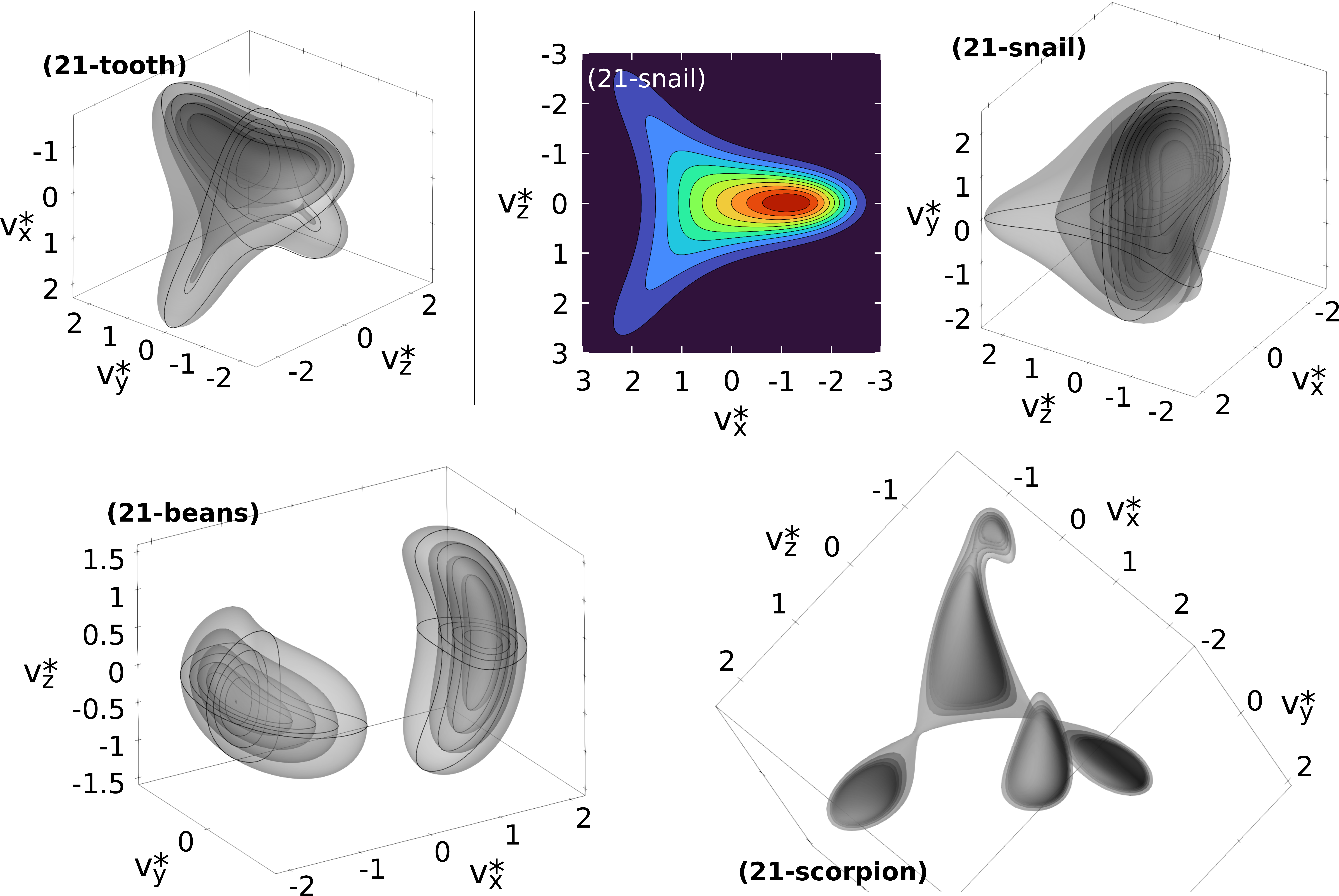}
\caption{Further shapes obtained from the 21-moment ME VDF. The two-dimensional contourplot for the case (21-snail) represents the plane $v_y^\star = 0$.}
\label{fig:VDF21-various-shapes}
\end{figure}

We report in this section some further shapes, encountered during the exploration of moment space.
These VDFs are presented without a particular order, and we assign them an arbitrary label based on their appearance.
The ME VDFs shown in Fig.~\ref{fig:VDF21-various-shapes} correspond to the states
\begin{description}
  \item[\rm(21-tooth)] $P_{ij}^\star = \delta_{ij}$,  $Q_{xxx}^\star = -Q_{xyy}^\star = 0.5$, $R_{iijj}^\star = 15 \ $;
  \item[\rm(21-snail)] $P_{xx}^\star = P_{yy}^\star = 1.2$, $P_{zz}^\star = 0.6$,  $Q_{xzz}^\star = 0.6573$, $R_{iijj}^\star = 10 \ $;
  \item[\rm(21-beans)] $P_{xx}^\star = 2$, $P_{yy}^\star = P_{zz}^\star = 0.5$,  $Q_{xzz}^\star = -Q_{xyy}^\star = 0.25$, $R_{iijj}^\star = 10 \ $;
  \item[\rm(21-scorpion)] $P_{xx}^\star = P_{zz}^\star = 1/2$, $P_{yy}^\star = 2$,  $Q_{ijk}^\star$ from Eq.~\eqref{eq:Qijk-headset-VDF}, $R_{iijj}^\star = 10 \ $;
\end{description}

\noindent where the heat flux tensor for the case (21-scorpion) reads
\begin{equation}\label{eq:Qijk-headset-VDF}
  \begin{cases}
    Q_{ijk}^\star = \sqrt{2}  \ \ \mathrm{if} \ \ i = j = k \, ,\\
    Q_{ijk}^\star = -\sqrt{2}/2 \ \ \mathrm{if} \ \ i \neq j = k \, , \\
    Q_{xyz}^\star = 0 \, .
  \end{cases}
\end{equation}

\section{Wave speeds of the maximum-entropy system}\label{sec:eigenvalues}

Most of the present interest in moment methods lies in their capability to offer a 
simplified alternative to kinetic formulations such as the Boltzmann or the Vlasov equations.
By computing moments of the full kinetic equation, and by integrating it over the whole velocity space, 
one can obtain a system of evolution equations for a finite set of moments of interest, $\bm{U}$.
This system can be written in balance-law form, as \cite{ferziger1972mathematical}
\begin{equation}\label{eq:PDE-cons-form}
  \frac{\partial \bm{U}}{\partial t} 
+ \frac{\partial \bm{F}_x}{\partial x} 
+ \frac{\partial \bm{F}_y}{\partial y} 
+ \frac{\partial \bm{F}_z}{\partial z} 
= 
  \bm{S} \, ,
\end{equation}

\noindent where $t$ is the time and ($x,y,z$) are the spacial coordinates, 
$\bm{F}_i$ is the vector of convected fluxes along the $i$-th direction, and $\bm{S}$ is a source-term vector
that typically includes the effect of collisions and external forces.
It can be shown that, as one closes the convected fluxes using the maximum-entropy approximation, 
the resulting system of equations is hyperbolic, as the flux Jacobian has real-valued eigenvalues \cite{levermore1996moment}.
We refer to these eigenvalues as the wave speeds of the system.

A study of the wave speeds gives an indication of both the speed and the direction at which information propagates through
the domain.
Equilibrium models, such as the Euler equations of gas dynamics, can only reproduce isotropic propagation, in the frame 
of the flow velocity.
On the other hand, non-equilibrium methods, such as the ME moment method, can instead predict a more complex scenario. 
Perturbations are expected to propagate faster in the spacial directions associated with a larger temperature (or pressure), 
and can be affected by the higher-order moments as well.

For a selected moment state, either at or out of equilibrium, the wave speeds of a moment system
are tightly connected to the shape of the assumed distribution function.
For instance, for the simple case of particles evolving along a single spacial direction (one-dimensional physics), 
the wave speeds of the fourth-order maximum-entropy system have been previously shown to be 
the optimal Gauss quadrature points of the ME distribution function \cite{mcdonald2013towards}.
In this section, we analyze the wave speeds of the 14 and 21-moment methods, for selected moment states.


\subsubsection*{Numerical procedure}\label{sec:numerical-procedure-eigenvalues}

First, one needs to obtain the flux Jacobian associated with the desired moment state.
This can be done, numerically, in a variety of ways, including algorithmic differentiation \cite{giroux2021approximation} 
and finite-difference approximations.
Here, we exploit instead a mathematical property of the maximum-entropy formulation, discussed in \cite{levermore1996moment,mcdonald2013affordable},
which allows us to write the flux Jacobian associated with the fluxes in the $x$ direction as
\begin{equation}\label{eq:Jacobian-maxent-potentials-integrals}
  \frac{\partial \bm{F}_x}{\partial \bm{U}} 
= \left< m \, v_x \bm{\Phi} \bm{\Phi}^{\intercal} e^{\bm{\alpha}^\intercal \bm{\Phi}}\right> \left[ \left< m \, \bm{\Phi} \bm{\Phi}^{\intercal} e^{\bm{\alpha}^\intercal \bm{\Phi}}\right> \right]^{-1}\, ,
\end{equation}

\noindent where $\bm{\Phi}$ is defined in Eq.~\eqref{eq:m-fourth-order-14-21} for the 14- and 21-moment systems.
The procedure goes as follows.
For the desired gas state, $\bm{U}$, one needs to compute the associated ME VDF, via Newton iterations.
This gives the vector of coefficients, $\bm{\alpha}$.
Then, one can integrate Eq.~\eqref{eq:Jacobian-maxent-potentials-integrals}, numerically.
Notice that this step requires one to build two 14-by-14 or 21-by-21 matrices, integrating its individual components.
Finally, the eigenvalues (wave speeds) of the flux Jacobian are computed numerically.


\subsubsection*{Selected results} 

We consider here the eigenvalues associated with three arbitrarily selected gas states:
a Maxwellian, case (14d) and the $x$-aligned version of case (14h-1), which we denote here as (14h-1x),
\begin{description}
  \item[\rm(14h-1x)] $P_{yy}^\star=P_{zz}^\star=P_{xx}^\star/50 \ , \ Q_{xjj}^\star = 2   \ , \ Q_{yjj}^\star = Q_{zjj}^\star = 0\ , \ R_{iijj}^\star = 15 \ $.
\end{description}

The eigenvalues are computed for both the 14 and the 21-moment methods.
In order to obtain the 21-moment VDF for the said cases, we proceed as follows:  
\begin{itemize}
  \item First, we compute the 14-moment ME VDFs for these cases;
  \item Then, from these VDFs, we numerically compute all the elements of the heat flux tensor, and utilize these values as an input
        for obtaining the 21-moment VDFs.
\end{itemize}

\begin{figure}[h]%
\centering
\includegraphics[width=0.9\textwidth]{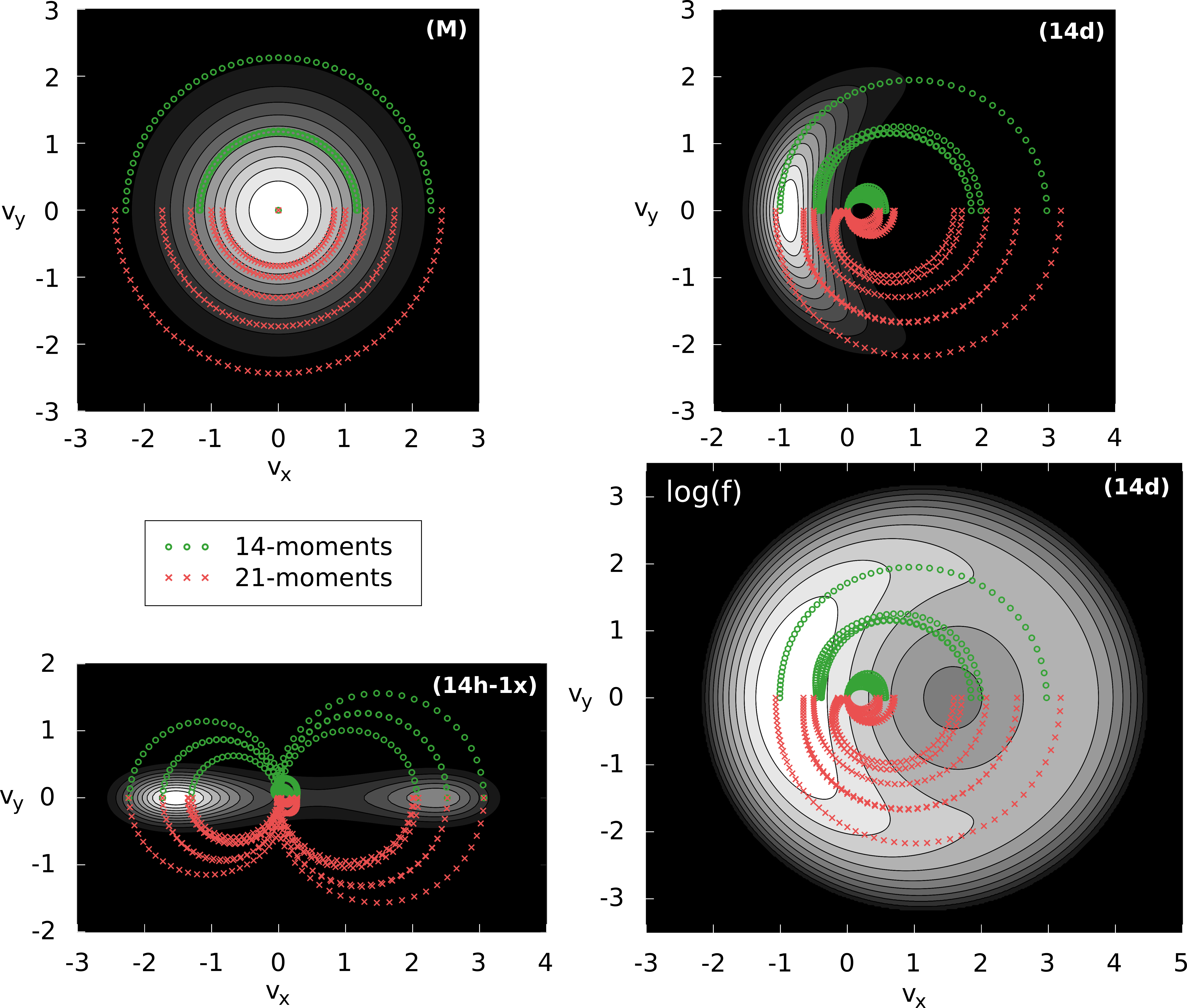}
\caption{Locus of the eigenvalues of the maximum-entropy moment systems, associated with the $x$ fluxes, at different angles,
         superimposed on the respective VDFs.
         \textbf{(M)} denotes the Maxwellian distribution. 
         The eigenvalue loci are symmetric with respect to $v_y$, and only half of the plot is shown:
         the upper half of each plot represents the 14-moment eigenvalues (green circles), while the lower half shows the
         21-moment ones (red crosses). }
\label{fig:eigenvalues-14-21-mom}
\end{figure}

\noindent In this way, the 14- and 21-moment methods produce the same VDFs, within numerical error.
Two-dimensional maps of the wave speeds are shown in Fig.~\ref{fig:eigenvalues-14-21-mom}, and are obtained as follows:
instead of computing and combining the $x$ and $y$ components of the wave speeds, obtained from the respective $\bm{F}_x$ and $\bm{F}_y$ fluxes, 
we exclusively compute the wave speeds along the $x$ direction, and repeat the calculation $50$ times by gradually rotating the reference frame.
After each computation, the resulting wave speeds are plotted at the corresponding angle.

Symmetric states, such as the Maxwellian VDF, result in perturbations propagating isotropically in space.
Instead, as shown in Figure~\ref{fig:eigenvalues-14-21-mom}, non-equilibrium VDFs are associated with anisotropic and/or asymmetric 
propagation.
For instance, elongated VDFs, such as that of case (14h-1x), show a preferential propagation in the direction of the elongation,
rather than towards its sides.
This is the continuum counterpart of the obvious kinetic interpretation that the direction of elongation contains faster particles.

Fig.~\ref{fig:eigenvalues-14-21-mom}-Top-Right and Bottom-Right show the same case, (14d).
From the Top-Right panel, a large number of the wave speeds are located in regions of the positive $v_x$ axis, where the VDF appears to be zero.
This might seem surprising, since a zero distribution function means that the considered region does not have any particle to support the propagation of
information.
However, this contradiction is merely apparent, and is easily solved by considering a logarithmic plot of the VDF (Bottom-Right panel) that
reveals a ring-like shape, although rather dim, and the presence of a significant number of particles.

For all cases, the maximum and minimum wave speeds of the 21-moment system are slightly larger than their 14-moment counterparts. 
However, the deviation is only marginal.
Future suggested studies include the analysis of the effect of individual elements of the $Q_{ijk}$ tensor on the wave speeds.
This will be particularly useful for the development of approximated formulas for the eigenvalues of the 21-moment system \cite{giroux2021approximation},
that are of large importance when strong non-equilibrium simulations are attempted \cite{boccelli2024numerical}.


\section{Conclusions}

In this work, we give an overview of (some of) the possible shapes assumed by the 14- and 
21-moment maximum-entropy (ME) velocity distribution functions (VDF), and discuss the associated moment states.
The ME VDFs are obtained numerically by employing Newton iterations.
The 14-moment maximum-entropy method is observed to reproduce toroidal, asymmetric, anisotropic and bi-modal distributions.
As the Junk subspace is approached, the VDFs are seen to split in a Maxwellian bulk, and a fast-moving 
low-density tail.
The 21-moment maximum-entropy VDF is able to reproduce every 14-moment distribution, and, within the limits imposed by 
the realizability contstraint, can further represent states with arbitrary entries in the heat flux tensor.
While the 14-moment method can produce VDFs with one or two local maxima, the 21-moment method can produce 
distributions with up to five peaks.
This model can also reproduce tetrahedral VDFs, as well as various additional shapes.

An analysis of the eigenvalues of the moment systems associated with the considered VDFs suggests that, 
as one might expect, the propagation of information in the maximum-entropy PDE systems roughly follows the particle 
paths identified by the (maximum-entropy) distribution functions.
Anisotropic states result in an anisotropic propagation of information, and higher-order moments also play a significant role.
In the conditions analyzed in this work, the 21-moment system reproduces slightly faster eigenvalues than the 
14-moment one.
Additional investigations of the 21-moment VDFs close to the realizability boundaries are suggested as a future work.

By illustrating the maximum-entropy VDFs associated with different gas states, the present study can provide the reader with
an intuitive understanding of which classes of problems can be profitably tackled by the 14- and 21-moment maximum-entropy methods.
However, it is important to mention that a perfect match of the distribution functions is not a strictly necessary
condition for the success of a moment method.
The actual final accuracy is better discussed on a case-by-case scenario.


\FloatBarrier
\appendix

\section{Computation of the energy distribution function}\label{sec:appendix-EDF}

The energy distribution function (EDF) is easily found by numerical integration of the VDF, 
along shells of constant energy.
First, one adopts a spherical representation of the velocity space, defining
\begin{equation}
    v_x = v \sin \theta \, \cos \phi \ \  , \ \ \  
    v_y = v \sin \theta \, \sin \phi \ \  ,  \ \ \ 
    v_z = v \cos \theta \, ,
\end{equation}

\noindent with $\theta \in [0, \pi]$ and $\phi \in [0, 2\pi)$.
In this system, 
\begin{equation}
  f(v_x, v_y, v_z) \, \mathrm{d}^3 v = v^2 \, \sin \theta \, f(v, \theta, \phi) \, \mathrm{d} v \, \mathrm{d} \theta \, \mathrm{d} \phi \, ,
\end{equation}

\noindent with $v = \| \bm{v} \|$. 
The distribution of velocity magnitudes, $f(v)$, is obtained by integrating over $\theta$ and $\phi$.
\noindent The EDF, $f(E)$, follows the equivalence relation
\begin{equation} 
  f(E) \, \mathrm{d} E = f(v) \, \mathrm{d} v \, ,
\end{equation} 

\noindent where, for a classical gas, $\mathrm{d} E = m v \, \mathrm{d} v$, where $m$ is the particle mass.
Ultimately, the EDF can be written as
\begin{equation}\label{eq:integral-fE}
  f(E) = \frac{v}{m} \, \int_{0}^{\pi} \int_{0}^{2\pi} \sin \theta \, f(v, \theta, \phi) \, \mathrm{d} \theta \, \mathrm{d} \phi \, .
\end{equation}








\end{document}